\documentclass[%
 reprint,
 superscriptaddress,
 amsmath,
 aps,
 prb
]{revtex4-2}

\usepackage[pdftex]{graphicx, color}
\usepackage[pdftex,
            colorlinks=true,
            citecolor=blue,
            linkcolor=blue]{hyperref}
\usepackage{newtxtext, newtxmath}
\usepackage{bm}
\usepackage{braket, mathtools}
\usepackage{bbm}
\usepackage{tabularx}
\graphicspath{{./images/}}

\newcommand{\ketbra}[2]{\ket{#1}\!\bra{#2}}
\DeclareMathOperator{\sgn}{sgn}

\begin{document}

\title{Superconductor/normal-metal/superconductor junction of topological superconductors revisited:\texorpdfstring{\\}{} Fractional Josephson current, fermion parity, and oscillating wavefunctions}

\author{Shuntaro Sumita}
\email[]{shuntaro.sumita@riken.jp}
\affiliation{%
 Condensed Matter Theory Laboratory, RIKEN CPR, Wako, Saitama 351-0198, Japan
}%

\author{Akira Furusaki}
\affiliation{%
 Condensed Matter Theory Laboratory, RIKEN CPR, Wako, Saitama 351-0198, Japan
}%
\affiliation{%
 RIKEN Center for Emergent Matter Science, Wako, Saitama 351-0198, Japan
}%

\date{\today}

\begin{abstract}
 The fractional Josephson effect is known to be a characteristic phenomenon of topological Josephson junctions hosting Majorana zero modes (MZMs), where the Josephson current has a $4\pi$ (rather than a $2\pi$) periodicity in the phase difference between the two topological superconductors.
 We introduce a one-dimensional model of a topological superconductor/normal-metal/superconductor (SNS) junction with the normal-metal (N) region of finite length, which is intermediate regime between the short- and long-junction limits.
 Assuming weak tunneling at the SN interfaces, we investigate resonance and finite-size effects on the fractional Josephson effect due to the existence of several discrete energy levels in the N region in which wavefunctions have oscillating nodal structure.
 Through careful analysis of the sign change in the transmission amplitudes through the junction and the fermion parity of the two MZMs, we find that the fractional Josephson current is proportional to the parity of total fermion numbers including both filled normal levels and two MZMs.
 Furthermore, we elucidate drastic enhancement of the Josephson current due to the resonance between a discrete level in the N region and MZMs.
\end{abstract}

\maketitle

\section{Introduction}
Topological phases of matter have recently attracted much attention in condensed matter physics~\cite{Hasan2010_review, Qi2011_review, Chiu2016_review}.
One of the characteristic features of topological phases of matter is the presence of symmetry-protected gapless boundary states that are robust against weak perturbations respecting relevant symmetry.
The robustness is related to topology of wavefunctions and is important from the perspective of device applications as well as fundamental physics.
Indeed, Majorana zero modes (MZMs) in a topological superconductor have been expected to be an essential element for quantum memory and computing~\cite{Kitaev2001, Nayak2008_review, Beenakker2013, Elliott2015_review, Sau2021_arXiv}.
Although many experimental studies have reported signatures of Majorana fermions so far~\cite{Rokhinson2012, Deng2012, Das2012, Churchill2013, Finck2013, Albrecht2016}, those results are inconclusive and nothing more than necessary conditions for the existence of MZMs~\cite{Chiu2019}, and a decisive evidence of their existence has not been obtained yet.

As a promising experimental proof for the existence of MZMs, previous theoretical studies have suggested a fractional Josephson effect in a topological Josephson junction, where the Josephson current has a $4\pi$, rather than a $2\pi$, periodicity as a function of the phase difference between the two topological superconductors~\cite{Kitaev2001, Kwon2004, Fu2009, Lutchyn2010, KTLaw2011, Nogueira2012, Pikulin2012, Beenakker2013_PRL, Zhang2014_PRB, Affleck2014, Sato2016, Sau2017, Alicea2012, Tanaka1996, Asano2006, Asano2013, Ikegaya2016}.
The first paper presenting this idea is Ref.~\cite{Kitaev2001}, in which Kitaev introduced a one-dimensional (1D) lattice model of a spinless $p$-wave superconductor (the so-called Kitaev chain), which is a prototypical model of a topological superconductor.
He showed that the ground-state fermion parity (FP) of a topological Josephson junction switches its sign every $2\pi$ change of the phase difference, resulting in the $4\pi$-periodic Josephson current.
There have been several experimental reports on the observation of the fractional Josephson effect~\cite{Rokhinson2012, Wiedenmann2016, Bocquillon2016, Li2018, Yu2018, Wang2018}.
More recently, it has been shown that many-body interactions or impurities can induce an $8\pi$-periodic fractional Josephson effect, due to the presence of $\mathbb{Z}_4$ parafermions or fractional MZMs~\cite{Zhang2014_PRL, Klinovaja2014, Orth2015, Vinkler-Aviv2017, Pedder2017, Shen2021}.

Regarding 1D topological superconductors themselves, recent studies have pointed out that the energy splitting of MZMs in a topological superconductor of finite length is, under certain conditions, an oscillating function of system parameters, such as a chemical potential, a superconducting gap, and a system size~\cite{Cheng2009, Cheng2010, Prada2012, Sarma2012, Pientka2013, Rainis2013, Ben-Shach2015, Thakurathi2015, Kao2014, Hegde2015, Hegde2016}.
In particular, Hegde and his collaborators have investigated a finite-size Kitaev chain, and elucidated that the number of times the ground-state FP switches depends on the chain length, inside the parameter region which they call the circle of oscillations~\cite{Hegde2015, Hegde2016}.
The switching behavior is attributed to oscillations in the Majorana wavefunction.

Similarly, we expect that a strong finite-size effect may exist in the fractional Josephson effect of topological superconductor/normal-metal/superconductor (SNS) junctions that have a MZM at each SN interface, for the following reasons.
First, the number of nodes in the wavefunction of the $n$th lowest energy level in the normal-metal (N) region depends on $n$.
In other words, the number of oscillations is controlled by the chemical potential.
Second, the fractional Josephson current is proportional to the \textit{first} power of the transmission amplitude of a single electron (rather than that of a Cooper pair as in conventional Josephson junctions) between the two superconductors, which necessarily involves the wavefunction of the Fermi level in the N region~\cite{Kitaev2001, Kwon2004, Fu2009}.
Third, when an energy level in the N region coincides with the Fermi energy of the superconductors (i.e., zero energy), the supercurrent is expected to be strongly enhanced due to resonance.
These considerations lead us to speculate that the sign (direction) of the fractional Josephson current can be switched by changing the chemical potential, due to the oscillating behavior of the normal wavefunctions.
However, to our knowledge, most of the previous studies have either considered the short-junction limit where there is no normal discrete levels or paid little attention to the oscillation in transmission amplitudes~\cite{Kwon2004, Fu2009, Lutchyn2010, KTLaw2011, Nogueira2012, Pikulin2012, Beenakker2013_PRL, Zhang2014_PRB, Affleck2014, Sato2016, Sau2017, Alicea2012, Tanaka1996, Asano2006, Asano2013, Ikegaya2016, Setiawan2017_1, Setiawan2017_2}.

With the above backgrounds, in this paper we revisit the problem of a 1D topological SNS junction.
We consider a lattice model of an SNS junction in which superconducting electrodes are represented by two semi-infinite Kitaev chains that are weakly connected through the N region of $L$ sites.
The transmission through SN interfaces is assumed to be very small, so that our model actually corresponds to an SINIS junction (I: insulator); we will nevertheless call it an SNS junction for simplicity throughout this paper.
Importantly, the total FP is conserved in our model, as our SNS junction is not coupled to electron reservoirs.

We focus on the intermediate-$L$ regime where there are several discrete energy levels in the N region, in which case the finite-size effects of our interest should be pronounced.
Then we calculate the low-energy spectrum of the model using a perturbation theory and a recently developed exact diagonalization method of a \textit{corner-modified banded block-Toeplitz matrix}~\cite{Alase2016, Cobanera2017, Alase2017, Cobanera2018}.
The energy splitting of the MZMs obtained from these calculations gives a direct measure of the fractional Josephson effect.
Consequently, we confirm resonant enhancement of supercurrent to occur every time a discrete level in the N region and MZMs are in resonance.
However, on the contrary to our naive expectation, we find that the direction of the supercurrent flow is not reversed by varying chemical potential in the N region.
The latter conclusion is reached by careful analysis of the interplay between the sign change in transmission amplitudes (or normal wavefunctions) and the FP of the two MZMs.
As a result we find that the fractional Josephson current is proportional to the parity of total fermion numbers including both filled normal levels and two MZMs.

The paper is organized as follows.
In Sec.~\ref{sec:model} our lattice model of the SNS junction with Kitaev chains as S electrodes is presented.
In Sec.~\ref{sec:perturbation} we obtain the energy spectrum of the model using second-order and first-order perturbations for off- and on-resonance cases, respectively.
Next, an effective model interpolating the two cases is constructed and analyzed with particular attention paid to the total and partial FPs in Sec.~\ref{sec:effective_model}.
In Sec.~\ref{sec:exact_diagonalization} we show results of the (numerical) exact diagonalization method indicating resonant enhancements of the fractional Josephson effect, in good agreement with the perturbation theory.
Finally, a brief summary and discussion are given in Sec.~\ref{sec:summary}.

\section{Model}
\label{sec:model}
\begin{figure}[tbp]
 \centering
 \includegraphics[width=\linewidth, pagebox=artbox]{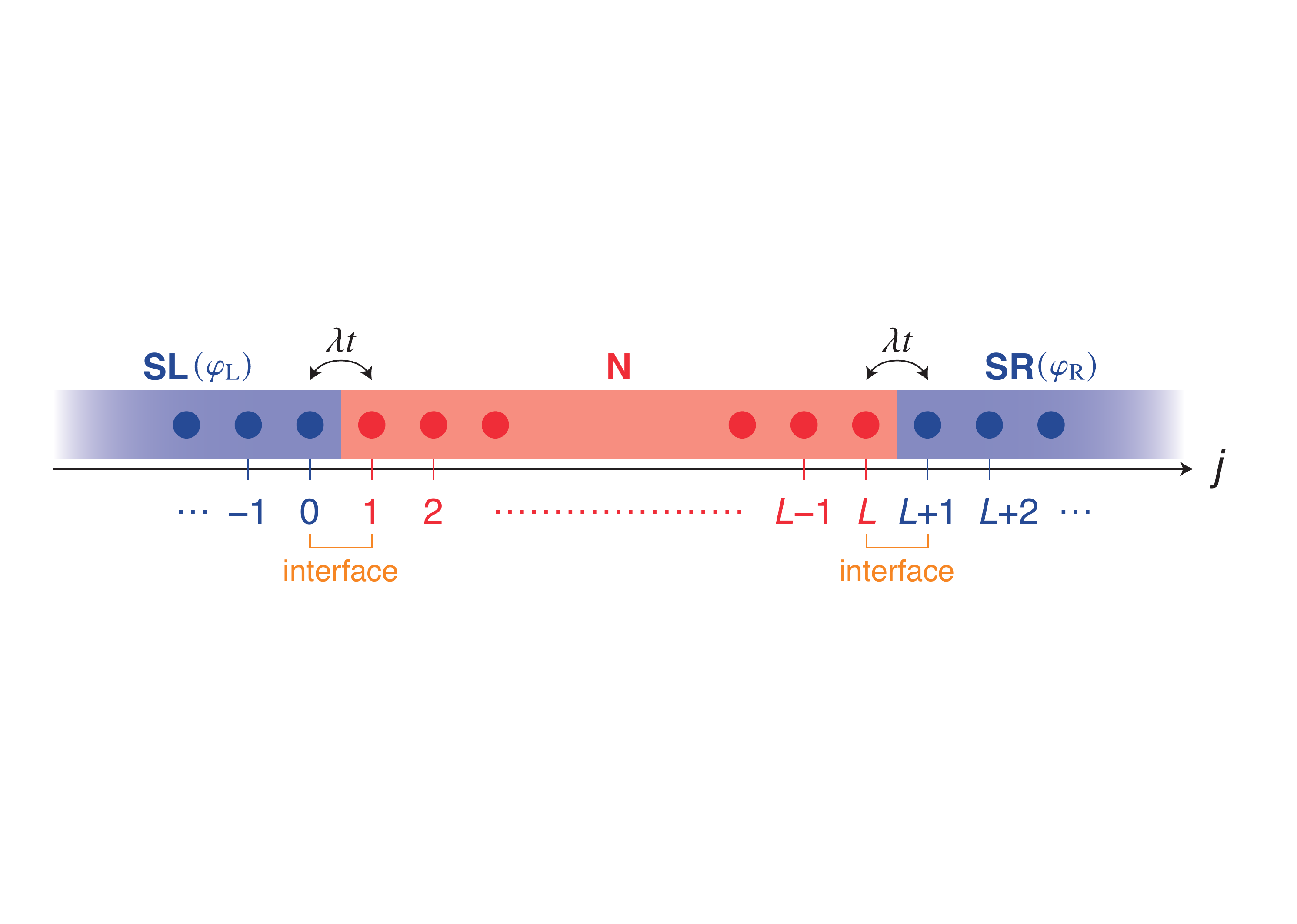}
 \caption{A schematic illustration of the 1D SNS junction described by the Hamiltonian~\eqref{eq:Hamiltonian}. The left (SL) and right (SR) superconducting regions are semi-infinite Kitaev chains with U(1) phases $\varphi_{\text{L}}$ and $\varphi_{\text{R}}$, respectively. The normal (N) region has finite $L$ sites.}
 \label{fig:SNS}
\end{figure}

First, we introduce a 1D simple tight-binding model describing an SNS junction with both of the superconductors being semi-infinite Kitaev chains,
\begin{align}
 H &= H_{\text{SL}} + H_{\text{N}} + H_{\text{SR}} + H_{\text{T}},
 \label{eq:Hamiltonian} \displaybreak[2] \\
 H_{\text{SL}} &= - 2 \sum_{j = -\infty}^{0} \mu \left(c_j^\dagger c_j - \frac{1}{2}\right) \notag \\
 & \quad - \sum_{j = -\infty}^{-1} (t c_j^\dagger c_{j+1} - \Delta e^{i\varphi_{\text{L}}} c_j^\dagger c_{j+1}^\dagger + \text{H.c.}),
 \label{eq:Hamiltonian_SL} \displaybreak[2] \\
 H_{\text{N}} &= - 2 \sum_{j = 1}^{L} \mu \left(c_j^\dagger c_j - \frac{1}{2}\right) - \sum_{j = 1}^{L-1} (t c_j^\dagger c_{j+1} + \text{H.c.}),
 \label{eq:Hamiltonian_N} \displaybreak[2] \\
 H_{\text{SR}} &= - 2 \sum_{j = L+1}^{\infty} \mu \left(c_j^\dagger c_j - \frac{1}{2}\right) \notag \\
 & \quad - \sum_{j = L+1}^{\infty} (t c_j^\dagger c_{j+1} - \Delta e^{i\varphi_{\text{R}}} c_j^\dagger c_{j+1}^\dagger + \text{H.c.}),
 \label{eq:Hamiltonian_SR} \displaybreak[2] \\
 H_{\text{T}} &= - \lambda (t c_0^\dagger c_1 + t c_L^\dagger c_{L+1} + \text{H.c.}),
 \label{eq:Hamiltonian_T}
\end{align}
where $c_j$ is a fermionic annihilation operator on $j$th site.
$t$ and $\mu$ represent a hopping parameter and a chemical potential, respectively, while $\Delta$ is an amplitude of the $p$-wave superconducting order parameter.
$H_{\text{SL}}$ and $H_{\text{SR}}$ represent semi-infinite Kitaev chains with superconducting U(1) phases $\varphi_{\text{L}}$ and $\varphi_{\text{R}}$, respectively, which are depicted by the blue regions in Fig.~\ref{fig:SNS}.
On the other hand, a normal metal in the middle of the junction (the red region in Fig.~\ref{fig:SNS}) is described by $H_{\text{N}}$ with finite $L$ sites (unit cells).
Furthermore, the Hamiltonian $H_{\text{T}}$ describes the tunneling across the SN interfaces with the hopping amplitude $\lambda t$.
For simplicity, $t, \Delta > 0$ and $0 \leq \lambda < 1$ are assumed throughout the paper.

For the later discussions, let us rewrite the second-quantized Hamiltonian [Eqs.~\eqref{eq:Hamiltonian_SL}--\eqref{eq:Hamiltonian_T}] into a first-quantized form.
The block-structure of the Hamiltonian enables us to decompose $H$ into lattice and internal state (particle-hole) spaces~\cite{Alase2016, Cobanera2017, Alase2017, Cobanera2018}.
Let $\{\ket{j} \mid j \in \mathbb{Z}\}$ be orthonormal bases of the lattice space; $\ket{j}$ corresponds to $j$th site.
Then the first-quantized Hamiltonian is given by Eq.~\eqref{eq:Hamiltonian} and
\begin{align}
 H_{\text{SL}} &= \mathbbm{1}_{\text{SL}} \otimes h_0^{\text{S}} + \left( T_{\text{SL}} \otimes h_1^{\text{S}}(\varphi_{\text{L}}) + T_{\text{SL}}^\dagger \otimes h_1^{\text{S}}(\varphi_{\text{L}})^\dagger \right),
 \label{eq:Hamiltonian_SL_1st} \\
 H_{\text{N}} &= \mathbbm{1}_{\text{N}} \otimes h_0^{\text{N}} + \left( T_{\text{N}} \otimes h_1^{\text{N}} + T_{\text{N}}^\dagger \otimes h_1^{\text{N} \dagger} \right),
 \label{eq:Hamiltonian_N_1st} \\
 H_{\text{SR}} &= \mathbbm{1}_{\text{SR}} \otimes h_0^{\text{S}} + \left( T_{\text{SR}} \otimes h_1^{\text{S}}(\varphi_{\text{R}}) + T_{\text{SR}}^\dagger \otimes h_1^{\text{S}}(\varphi_{\text{R}})^\dagger \right),
 \label{eq:Hamiltonian_SR_1st} \\
 H_{\text{T}} &= \lambda \left[ (\ketbra{0}{1} + \ketbra{L}{L+1}) \otimes h_1^{\text{N}} + \text{H.c.} \right],
 \label{eq:Hamiltonian_T_1st}
\end{align}
where the identity and left-shift operators for each region are defined as
\begin{subequations}
 \begin{align}
  \mathbbm{1}_{\text{SL}} &= \sum_{j=-\infty}^{0} \ketbra{j}{j}, & T_{\text{SL}} &= \sum_{j=-\infty}^{-1} \ketbra{j}{j+1}, \\
  \mathbbm{1}_{\text{N}} &= \sum_{j=1}^{L} \ketbra{j}{j}, & T_{\text{N}} &= \sum_{j=1}^{L-1} \ketbra{j}{j+1}, \\
  \mathbbm{1}_{\text{SR}} &= \sum_{j=L+1}^{\infty} \ketbra{j}{j}, & T_{\text{SR}} &= \sum_{j=L+1}^{\infty} \ketbra{j}{j+1}.
 \end{align}
\end{subequations}
Furthermore, $h^{\text{S, L}}_r$ represents a matrix in the particle-hole space associated with $r$ shifts (hoppings) of sites:
\begin{subequations}
 \begin{align}
  h_0^{\text{S}} &= h_0^{\text{N}} =
  \begin{bmatrix}
   -\mu & 0 \\
   0 & \mu
  \end{bmatrix}, \\
  h_1^{\text{S}}(\varphi) &= \frac{1}{2}
  \begin{bmatrix}
   -t & \Delta e^{i\varphi} \\
   -\Delta e^{-i\varphi} & t
  \end{bmatrix}, \,
  h_1^{\text{N}} = \frac{1}{2}
  \begin{bmatrix}
   -t & 0 \\
   0 & t
  \end{bmatrix}.
 \end{align}
\end{subequations}

It is well known that a semi-infinite Kitaev chain possesses a MZM localized at its end, when the chemical potential satisfies $|\mu / t| < 1$.
Throughout this paper, we focus on this parameter regime since we are interested in the Josephson effect in the presence of MZMs on both SN interfaces.

\section{Perturbation theory}
\label{sec:perturbation}
In this section, we investigate the energy spectrum of the SNS junction Hamiltonian [Eq.~\eqref{eq:Hamiltonian}] in terms of a perturbation theory.
Assuming $\lambda \ll 1$, we treat the tunneling Hamiltonian $H_{\text{T}}$ as a perturbation to the Hamiltonian $H_{\text{SL}} + H_{\text{N}} + H_{\text{SR}}$.
The advantage of this approach is that the unperturbed Hamiltonian describes the S and N regions separately and is easy to solve (Sec.~\ref{sec:perturbation_preliminary}).
Next, we derive energy level structures when eigenstates of $H_{\text{N}}$ are off (on) resonance with MZMs, by using a second-order (first-order) perturbation theory in Sec.~\ref{sec:perturbation_off-resonance} (Sec.~\ref{sec:perturbation_on-resonance}).

\subsection{Preliminary}
\label{sec:perturbation_preliminary}
First, we show energy eigenvalues and eigenstates of the unperturbed Hamiltonian $H_{\text{SL}} + H_{\text{N}} + H_{\text{SR}}$.
The detailed derivation is given in Appendix~\ref{app:exact_sol}.

Since we assume $|\mu / t| < 1$, the semi-infinite left and right superconducting chains have MZMs localized at $j = 0$ and $j = L+1$, respectively~%
\footnote{The eigenstates of $H_{\text{SL}}$ and $H_{\text{SR}}$ outside the superconducting gap are not considered in this study, because the interaction of the MZMs is our main concern.}.
The zero-energy eigenstates of $H_{\text{SL}}$ and $H_{\text{SR}}$ are given by
\begin{align}
 \Ket{\psi^{\text{SL}}} &= N^{\text{S}} \sum_{j=-\infty}^{0} \left[ (\mathsf{z}^{\text{S}}_1)^{-j+1} - (\mathsf{z}^{\text{S}}_2)^{-j+1} \right] \ket{j}
 \begin{bmatrix}
  e^{i\varphi_{\text{L}}/2} \\ e^{-i\varphi_{\text{L}}/2}
 \end{bmatrix},
 \label{eq:Kitaev_SL} \\
 \Ket{\psi^{\text{SR}}} &= N^{\text{S}} \sum_{j=L+1}^{\infty} \left[ (\mathsf{z}^{\text{S}}_1)^{j-L} - (\mathsf{z}^{\text{S}}_2)^{j-L} \right] \ket{j}
 \begin{bmatrix}
  i e^{i\varphi_{\text{R}}/2} \\ -i e^{-i\varphi_{\text{R}}/2}
 \end{bmatrix},
 \label{eq:Kitaev_SR}
\end{align}
where $N^{\text{S}}$ is a normalization constant [see Eq.~\eqref{eq:norm_S}], and
\begin{equation}
 \mathsf{z}^{\text{S}}_{1, 2} = \frac{- \mu \pm \sqrt{\mu^2 + \Delta^2 - t^2}}{t + \Delta}.
\end{equation}
Equations~\eqref{eq:Kitaev_SL} and \eqref{eq:Kitaev_SR} are derived under the assumption that $\mu^2 + \Delta^2 - t^2 \neq 0$, that is, $\mathsf{z}^{\text{S}}_1 \neq \mathsf{z}^{\text{S}}_2$; see also Eqs.~\eqref{eq:Kitaev_power-law_SL} and \eqref{eq:Kitaev_power-law_SR} for zero-energy eigenvectors when $\mu^2 + \Delta^2 - t^2 = 0$.
Note that $\mathsf{z}^{\text{S}}_{1, 2}$ are complex numbers when $\mu^2 + \Delta^2 - t^2 < 0$.
The normal-metal Hamiltonian $H_{\text{N}}$ has $2L$ eigenstates in the bulk (and no end state), where the coefficient $2$ comes from the particle-hole sectors in our Bogoliubov--de Gennes formalism.
The eigenenergies and eigenstates are
\begin{gather}
 \varepsilon_{1, 2}^{\text{N}}(q) = \mp \left[\mu + t \cos\left(\frac{\pi q}{L + 1}\right)\right] =: \pm \varepsilon^{\text{N}}(q),
 \label{eq:energy_N} \\
 \Ket{\psi^{\text{N}}_{1, 2}(q)} = N^{\text{N}} \sum_{j=1}^{L} \sin\left(\frac{\pi q j}{L + 1}\right) \ket{j} \left\{
 \begin{bmatrix}
  1 \\ 0
 \end{bmatrix}, \,
 \begin{bmatrix}
  0 \\ 1
 \end{bmatrix}
 \right\},
\end{gather}
where $N^{\text{N}}$ is a normalization constant [see Eq.~\eqref{eq:norm_N}], and $q = 1, 2, \dots, L$.
$\ket{\psi^{\text{N}}_1(q)}$ and $\ket{\psi^{\text{N}}_2(q)}$ correspond to particle and hole wavefunctions, respectively; they have energies of opposite sign, $\pm \varepsilon^{\text{N}}(q)$.
Note that the $q$th particle/hole wavefunctions $\ket{\psi^{\text{N}}_l(q)}$ possess $q - 1$ sign changes between $j = 1$ and $j = L$.

In the following subsections, we discuss the perturbation theory with the small parameter $\lambda$ in the tunneling Hamiltonian $H_{\text{T}}$.
The tunneling matrix elements of $H_{\text{T}}$ between the superconducting and normal-metal wavefunctions are given by
\begin{subequations}
 \label{eq:HT_matrix_element}
 \begin{align}
  \Braket{\psi^{\text{N}}_l(q) | H_{\text{T}} | \psi^{\text{SL}}} &=
  \begin{cases}
   - \tilde{t}(q) e^{i\varphi_{\text{L}}/2}, & l = 1, \\
   + \tilde{t}(q) e^{-i\varphi_{\text{L}}/2}, & l = 2,
  \end{cases}
  \label{eq:HT_matrix_element_SL} \\
  \Braket{\psi^{\text{N}}_l(q) | H_{\text{T}} | \psi^{\text{SR}}} &=
  \begin{cases}
   i\tilde{t}(q) (-1)^q e^{i\varphi_{\text{R}}/2}, & l = 1, \\
   i\tilde{t}(q) (-1)^q e^{-i\varphi_{\text{R}}/2}, & l = 2,
  \end{cases}
  \label{eq:HT_matrix_element_SR}
 \end{align}
\end{subequations}
where
\begin{align}
 \tilde{t}(q) = N^{\text{S}} N^{\text{N}} \lambda t \frac{\sqrt{\mu^2 + \Delta^2 - t^2}}{t + \Delta} \sin\left(\frac{\pi q}{L + 1}\right).
\end{align}
The $(-1)^q$ factor in Eq.~\eqref{eq:HT_matrix_element_SR} reflects the fact that the wavefunctions in the N region have $q - 1$ nodes.

\begin{figure}[tbp]
 \centering
 \includegraphics[width=\linewidth, pagebox=artbox]{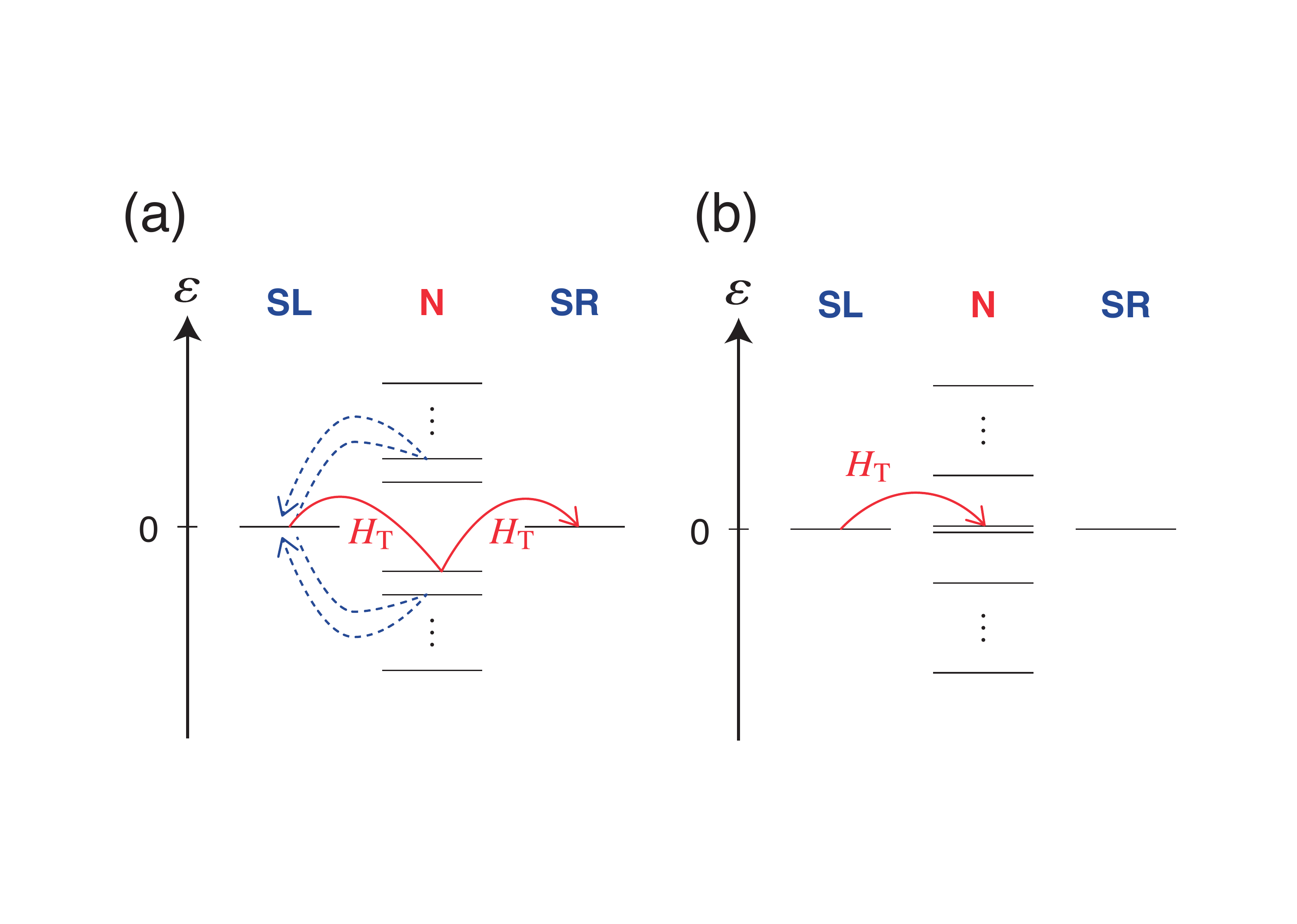}
 \caption{Schematic illustrations of single-particle energy levels (a) without and (b) with zero energy states in the N region. The lowest perturbation theory for the tunneling Hamiltonian $H_{\text{T}}$ is (a) second-order and (b) first-order, respectively. Some transition processes due to the perturbation are shown by the red and blue arrows.}
 \label{fig:energy_perturbation}
\end{figure}

When calculating energy splitting of the MZMs by the perturbation $H_{\text{T}}$, we need to consider two cases separately:
\begin{description}
 \item[off-resonance case] all of the normal energy eigenvalues in Eq.~\eqref{eq:energy_N} being away from zero energy [Fig.~\ref{fig:energy_perturbation}(a)], and
 \item[on-resonance case] two (one particle-like and one hole-like) of the normal energy eigenvalues being close to zero energy [Fig.~\ref{fig:energy_perturbation}(b)].
\end{description}
We will discuss the two cases in the following two subsections, respectively.

\subsection{Second-order perturbation in off-resonance case}
\label{sec:perturbation_off-resonance}
In this subsection, we discuss the off-resonance case, in which a second-order perturbation theory should be a good approximation as long as
\begin{equation}
 |\tilde{t}(q)| \ll |\varepsilon^{\text{N}}(q)|, \ \forall q = 1, \dots, L,
 \label{eq:2nd_perturbation_condition}
\end{equation}
is satisfied; see Fig.~\ref{fig:energy_perturbation}(a).

\begin{figure}[tbp]
 \includegraphics[width=\linewidth]{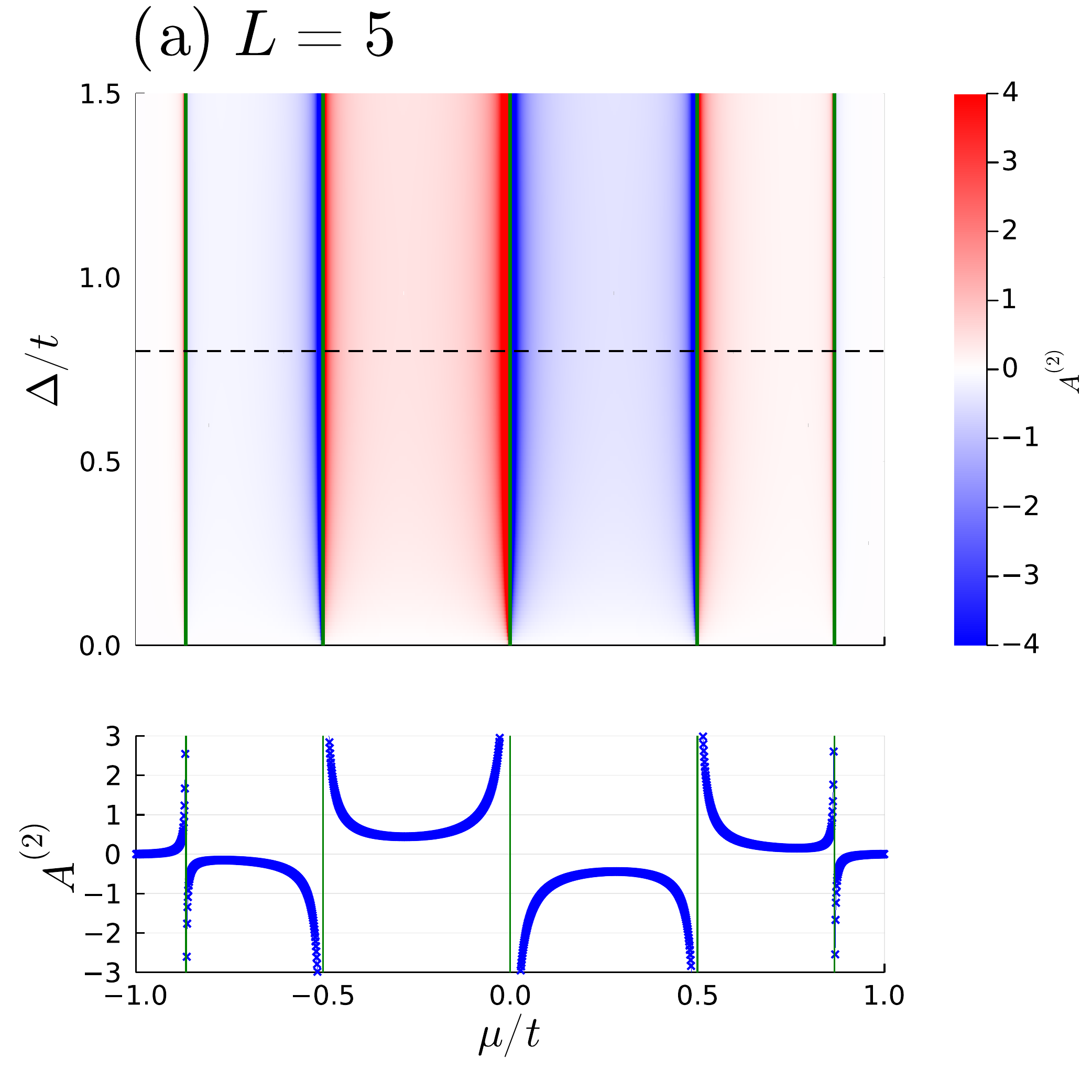}
 \includegraphics[width=\linewidth]{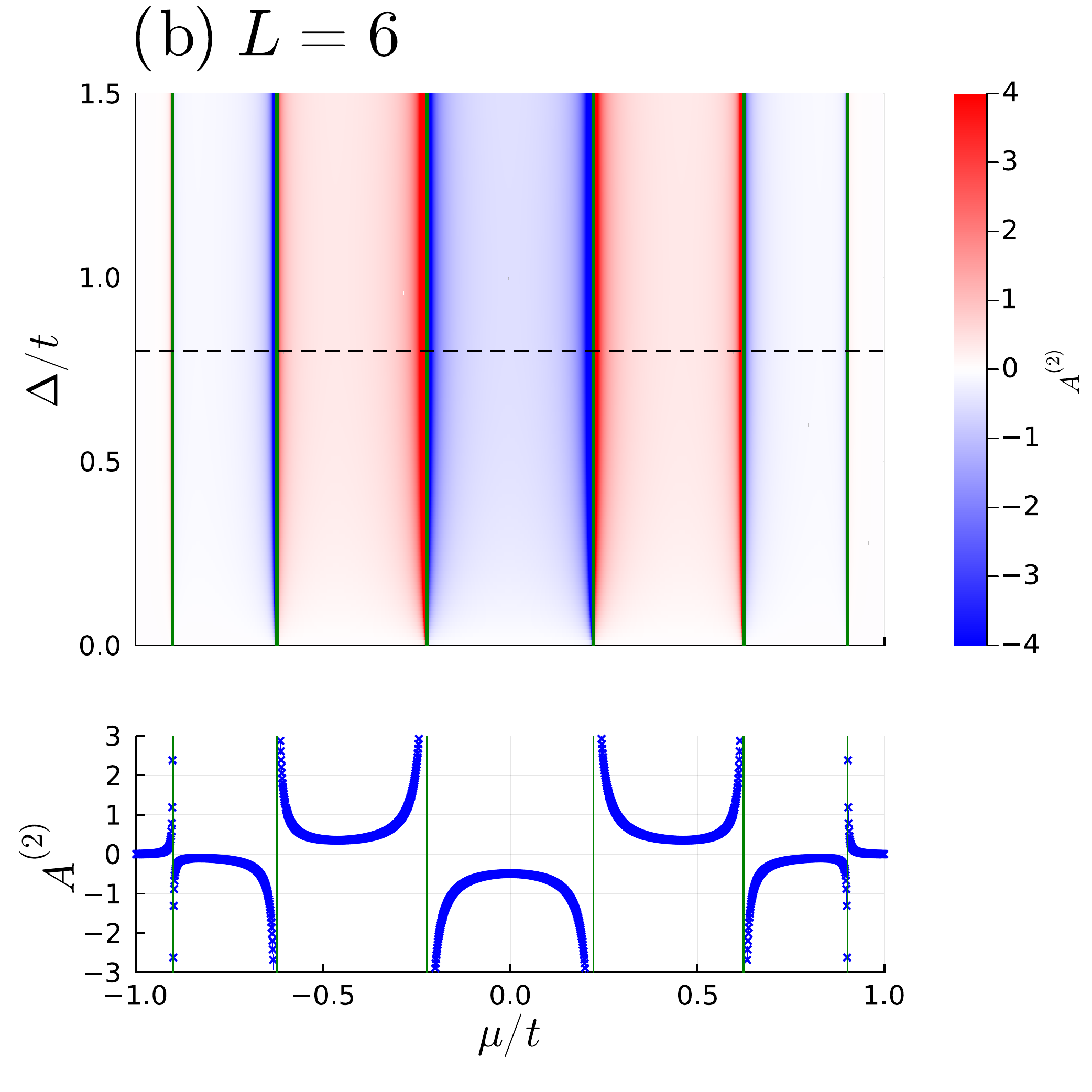}
 \caption{The parameter dependence of the ``amplitude'' function [Eq.~\eqref{eq:2nd_perturbed_amplitude}] in second-order perturbation theory for (a) $L = 5$ and (b) $L = 6$. The top panels in (a) and (b) are heat maps of $A^{(2)}$ in $\bar{\mu}$--$\bar{\Delta}$ plane, while the bottom panels show the $\bar{\mu}$ dependence for $\bar{\Delta} = 0.8$ (the black dashed lines in the top panels). The resonance points where $\varepsilon^{\text{N}}(q) = 0 \ (q = 1, \dots, L)$ is satisfied are illustrated by the green solid lines.}
 \label{fig:2nd_perturbed_amplitude}
\end{figure}

Since our main concern is the mixing of the two MZMs, initial states are fixed to be $\ket{\psi^{\text{SL}}}$ and $\ket{\psi^{\text{SR}}}$.
Introducing a projection operator $P_0 = \ketbra{\psi^{\text{SL}}}{\psi^{\text{SL}}} + \ketbra{\psi^{\text{SR}}}{\psi^{\text{SR}}}$, the second-order perturbed tunneling Hamiltonian is given by
\begin{equation}
 H_{\text{T}}^{(2)} = P_0 \sum_{q=1}^{L} \sum_{l=1}^{2} H_{\text{T}} \frac{\ketbra{\psi^{\text{N}}_l(q)}{\psi^{\text{N}}_l(q)}}{(-1)^l \varepsilon^{\text{N}}(q)} H_{\text{T}} P_0.
\end{equation}
Using the matrix elements in Eq.~\eqref{eq:HT_matrix_element}, the second-order Hamiltonian for the initial states $[\ket{\psi^{\text{SL}}}, \ket{\psi^{\text{SR}}}]$ is written in the matrix form
\begin{equation}
 \hat{H}_{\text{T}}^{(2)} = \lambda^2 t A^{(2)}(\bar{\mu}, \bar{\Delta}, L) \cos\left(\frac{\varphi_{\text{R}} - \varphi_{\text{L}}}{2}\right) \sigma_y,
 \label{eq:2nd_perturbed_Hamiltonian}
\end{equation}
where $\bar{\mu} := \mu / t$, $\bar{\Delta} := \Delta / t$, $\sigma_y$ is a Pauli matrix, and
\begin{align}
 A^{(2)}(\bar{\mu}, \bar{\Delta}, L) &= - \frac{1}{\lambda^2 t} \sum_{q=1}^{L} \frac{2(-1)^q |\tilde{t}(q)|^2}{\varepsilon^{\text{N}}(q)} \notag \\
 &= (N^{\text{S}} N^{\text{N}})^2 \frac{|\bar{\mu}^2 + \bar{\Delta}^2 - 1|}{(1 + \bar{\Delta})^2} \sum_{q=1}^{L} \frac{2(-1)^q \sin^2\left(\frac{\pi q}{L+1}\right)}{\bar{\mu} + \cos\left(\frac{\pi q}{L+1}\right)},
 \label{eq:2nd_perturbed_amplitude}
\end{align}
is a dimensionless ``amplitude'' function (see Fig.~\ref{fig:2nd_perturbed_amplitude}).
Note that the diagonal components of the matrix \eqref{eq:2nd_perturbed_Hamiltonian}, which correspond to to-and-fro procedures such as blue dashed arrows in Fig.~\ref{fig:energy_perturbation}(a), are zero because the contributions from the intermediate states $+ \varepsilon^{\text{N}}(q)$ and $- \varepsilon^{\text{N}}(q)$ cancel.

Away from resonance, we can obtain the energy splitting of the MZMs due to the tunneling by diagonalizing Eq.~\eqref{eq:2nd_perturbed_Hamiltonian},
\begin{equation}
 \varepsilon^{(2)}_\pm = \pm \lambda^2 t A^{(2)}(\bar{\mu}, \bar{\Delta}, L) \cos\left(\frac{\varphi_{\text{R}} - \varphi_{\text{L}}}{2}\right).
 \label{eq:2nd_perturbed_energy}
\end{equation}
Equation~\eqref{eq:2nd_perturbed_energy} indicates the existence of degeneracy at $\varphi_{\text{R}} - \varphi_{\text{L}} = (2n + 1)\pi \ (n \in \mathbb{Z})$ of the zero-energy states, which is protected by the FP (see also Sec.~\ref{sec:effective_model}).
The $\varphi_{\text{R}} - \varphi_{\text{L}}$ dependence of the ground-state energy (the lowest energy in both FP sectors) is given by
\begin{align}
 E_{\text{GS}}^{\text{off-res}} &\simeq \sum_{\sigma \in \text{occupied}} \varepsilon^{(2)}_\sigma + \text{const} \notag \\
 &= - \lambda^2 t \left| A^{(2)}(\bar{\mu}, \bar{\Delta}, L) \cos\left(\frac{\varphi_{\text{R}} - \varphi_{\text{L}}}{2}\right) \right| + \text{const},
 \label{eq:2nd_perturbed_energy_GS}
\end{align}
where const is independent of the relative phase.
The ground-state FP switches at $\varphi_{\text{R}} - \varphi_{\text{L}} = (2n + 1)\pi$, and for fixed FP we have $4\pi$ periodic (fractional) Josephson effect, as we will discuss in more detail in Sec.~\ref{sec:effective_model}~\cite{Kitaev2001, Kwon2004, Fu2009, Lutchyn2010, Nogueira2012, Zhang2014_PRB, Affleck2014}.

We here comment on the ``amplitude'' function $A^{(2)}(\bar{\mu}, \bar{\Delta}, L)$ in Eq.~\eqref{eq:2nd_perturbed_amplitude}.
Figures~\ref{fig:2nd_perturbed_amplitude}(a) and \ref{fig:2nd_perturbed_amplitude}(b) show the $(\bar{\mu}, \bar{\Delta})$ dependence of the function.
$A^{(2)}$ obviously changes its sign $L$ times as a function of the scaled chemical potential $\bar{\mu}$, reflecting the oscillating behavior of the wavefunctions in the N region.
Furthermore, the function diverges on the green lines defined by $\varepsilon^{\text{N}}(q) = 0 \ (q = 1, \dots, L)$, namely at the resonance points.
The divergent behavior of $A^{(2)}$ is not surprising, considering that Eq.~\eqref{eq:2nd_perturbation_condition} is not satisfied around the lines, where the second-order perturbation theory breaks down.
In the next subsection, we discuss a first-order perturbation theory that works for the on-resonance region.

\subsection{First-order perturbation in on-resonance case}
\label{sec:perturbation_on-resonance}
Let us move on to the on-resonance case, where there exists an integer $q_0$ ($1 \leq q_0 \leq L$) such that $\varepsilon^{\text{N}}(q_0) = 0$.
We here note that the case when the chemical potential $\mu = 0$ is on resonance for odd $L$, while it is off resonance for even $L$ (see also Fig.~\ref{fig:2nd_perturbed_amplitude}).
This fact is used for later discussions in Sec.~\ref{sec:exact_diagonalization_numerics}.

On the resonance point, four single-particle eigenstates $\ket{\psi^{\text{SL}}}$, $\ket{\psi^{\text{N}}_1(q_0)}$, $\ket{\psi^{\text{N}}_2(q_0)}$, and $\ket{\psi^{\text{SR}}}$ are degenerate (in resonance) at the zero energy [see Fig.~\ref{fig:energy_perturbation}(b)].
Therefore, a first-order perturbation theory can be applied to the analysis of energy level structures around the zero point.
Using Eq.~\eqref{eq:HT_matrix_element}, the first-order perturbation Hamiltonian matrix in the basis $[\ket{\psi^{\text{SL}}}, \ket{\psi^{\text{SR}}}, \ket{\psi^{\text{N}}_1(q_0)}, \ket{\psi^{\text{N}}_2(q_0)}]$ is given by
\begin{align}
 \hat{H}_{\text{T}}^{(1)} &= \lambda t A^{(1)}(q_0, \bar{\Delta}, L)
 \begin{bmatrix}
  0 & \hat{h}_{\text{T}}^{(1) \dagger} \\
  \hat{h}_{\text{T}}^{(1)} & 0
 \end{bmatrix},
 \label{eq:1st_perturbed_Hamiltonian} \\
 \hat{h}_{\text{T}}^{(1)} &= \frac{1}{2} s(q_0, \bar{\Delta})
 \begin{bmatrix}
  - e^{i\varphi_{\text{L}}/2} & i (-1)^{q_0} e^{i\varphi_{\text{R}}/2} \\
  e^{-i\varphi_{\text{L}}/2} & i (-1)^{q_0} e^{-i\varphi_{\text{R}}/2} \\
 \end{bmatrix},
\end{align}
where
\begin{align}
 A^{(1)}(q_0, \bar{\Delta}, L) &= \frac{2|\tilde{t}(q)|}{\lambda t} \notag \\
 &= 2 N^{\text{S}} N^{\text{N}} \frac{\left| \bar{\Delta}^2 - \sin^2\left(\frac{\pi q_0}{L+1}\right) \right|^{1/2}}{1 + \bar{\Delta}} \sin\left(\frac{\pi q_0}{L+1}\right), \\
 s(q_0, \bar{\Delta}) &=
 \begin{cases}
  1 & \text{for} \ |\bar{\Delta}| > \sin\left(\frac{\pi q_0}{L+1}\right), \\
  i & \text{for} \ |\bar{\Delta}| < \sin\left(\frac{\pi q_0}{L+1}\right).
 \end{cases}
\end{align}
Note that, when $\varepsilon^{\text{N}}(q_0) = 0$, the scaled chemical potential is determined by the integer $q_0$: $\bar{\mu} = - \cos\left(\frac{\pi q_0}{L+1}\right)$.

Diagonalizing the matrix \eqref{eq:1st_perturbed_Hamiltonian}, we obtain the energy splitting of the MZMs due to the tunneling:
\begin{equation}
 \varepsilon^{(1)}_{\sigma_1 \sigma_2} = \sigma_1 \lambda t A^{(1)}(q_0, \bar{\Delta}, L) \cos\left(\frac{\varphi_{\text{R}} - \varphi_{\text{L}} + \sigma_2 \pi}{4}\right),
 \label{eq:1st_perturbed_energy}
\end{equation}
with $\sigma_1, \sigma_2 \in \{\pm 1\}$.
Therefore, the dependence of the ground-state energy (the lowest energy of the two FP sectors) on the $\varphi_{\text{R}} - \varphi_{\text{L}}$ at resonance is given by
\begin{align}
 E_{\text{GS}}^{\text{on-res}} &\simeq \sum_{\{\sigma_1, \sigma_2\} \in \text{occupied}} \varepsilon^{(1)}_{\sigma_1 \sigma_2} + \text{const} \notag \\
 &= - \sqrt{2} \lambda t A^{(1)}(q_0, \bar{\Delta}, L) \notag \\
 &\qquad \times \max\left\{ \left|\cos\left(\frac{\varphi_{\text{R}} - \varphi_{\text{L}}}{4}\right)\right|, \left|\sin\left(\frac{\varphi_{\text{R}} - \varphi_{\text{L}}}{4}\right)\right| \right\} \notag \\
 & \quad + \text{const}.
 \label{eq:1st_perturbed_energy_GS}
\end{align}
From Eq.~\eqref{eq:1st_perturbed_energy_GS}, one may consider that the Josephson current has an $8\pi$ periodicity, rather than the $4\pi$ periodicity in Eq.~\eqref{eq:2nd_perturbed_energy_GS}.
As shown in the next section, however, the periodicity in $\varphi_{\text{R}} - \varphi_{\text{L}}$ of the lowest energy and Josephson current under fixed FP is $4\pi$ for both on- and off-resonance cases.

\section{Effective model connecting on-resonance and off-resonance cases}
\label{sec:effective_model}
In Sec.~\ref{sec:perturbation_off-resonance}, the energy level structures for the off-resonance case were discussed by using the second-order perturbation theory, which however breaks down around the resonance points.
Alternatively, the first-order perturbation theory is used for the on-resonance case (Sec.~\ref{sec:perturbation_on-resonance}).
In this section, we discuss an effective four-state model that can give a unified description of both on- and off-resonance situations.

\subsection{Effective model}
Let us consider the situation where the two (left and right) MZMs and the normal state labeled by $q = q_0$ are the four energy levels closest to zero energy in the superconducting gap.
An effective model focusing only on these states is thus constructed, which turns out to be equivalent to the one considered by Affleck and Giuliano~\cite{Affleck2014}.
The Affleck--Giuliano Hamiltonian is written in the notations used in the present paper as
\begin{align}
 H_{\text{AG}} &= \left(c^\dagger \Braket{\psi^{\text{N}}_1(q_0) | H_{\text{T}} | \psi^{\text{SL}}} \gamma_{\text{L}} + \text{H.c.}\right) \notag \\
 &\qquad + \left(c^\dagger \Braket{\psi^{\text{N}}_1(q_0) | H_{\text{T}} | \psi^{\text{SR}}} \gamma_{\text{R}} + \text{H.c.}\right) \notag \\
 &= - \tilde{t}_{\text{L}} e^{i\varphi_{\text{L}}/2} c^\dagger \gamma_{\text{L}} - \tilde{t}_{\text{L}}^* e^{-i\varphi_{\text{L}}/2} \gamma_{\text{L}} c \notag \\
 &\qquad - i \tilde{t}_{\text{R}} e^{i\varphi_{\text{R}}/2} c^\dagger \gamma_{\text{R}} + i \tilde{t}_{\text{R}}^* e^{-i\varphi_{\text{R}}/2} \gamma_{\text{R}} c,
\end{align}
where
\begin{equation}
 \tilde{t}_{\text{L}} = \tilde{t}(q_0), \quad \tilde{t}_{\text{R}} = (-1)^{q_0-1} \tilde{t}(q_0).
 \label{eq:effective_tL_tR}
\end{equation}
$c$ is an annihilation operator of the normal mode labeled by $q = q_0$, while $\gamma_{\text{L}}$ and $\gamma_{\text{R}}$ represent Majorana fermionic operators in the left and right superconducting regions, respectively.

The deviation from the resonance is controlled by a finite energy $\varepsilon^{\text{N}}$ of the normal particle.
Then we obtain the effective Hamiltonian
\begin{equation}
 H_{\text{eff}} = \varepsilon^{\text{N}} \left(c^\dagger c - \frac{1}{2}\right) + H_{\text{AG}},
\end{equation}
where the constant $\frac{1}{2}$ is introduced so that the energy spectrum of $H_{\text{eff}}$ is particle-hole symmetric.
The effective Hamiltonian is solved by introducing
\begin{equation}
 f = \frac{1}{2} (\gamma_{\text{L}} + i\gamma_{\text{R}}), \quad f^\dagger = \frac{1}{2} (\gamma_{\text{L}} - i\gamma_{\text{R}}),
\end{equation}
which, respectively, represent annihilation and creation operators of a nonlocal fermion produced from the two Majorana fermions.
Then the Hamiltonian is rewritten as
\begin{align}
 H_{\text{eff}} = \varepsilon^{\text{N}} \left(c^\dagger c - \frac{1}{2}\right)
 &- \left[ \left(\tilde{t}_{\text{L}} e^{i\varphi_{\text{L}}/2} - \tilde{t}_{\text{R}} e^{i\varphi_{\text{R}}/2}\right) c^\dagger f^\dagger + \text{H.c.} \right] \notag \\
 &- \left[ \left(\tilde{t}_{\text{L}} e^{i\varphi_{\text{L}}/2} + \tilde{t}_{\text{R}} e^{i\varphi_{\text{R}}/2}\right) c^\dagger f + \text{H.c.} \right].
 \label{eq:effective_Hamiltonian}
\end{align}
The Hamiltonian commutes with the operator
\begin{equation}
 (-1)^{F_{\text{eff}}} = e^{i\pi (c^\dagger c + f^\dagger f)} = (1 - 2c^\dagger c) (1 - 2f^\dagger f),
 \label{eq:partial_FP}
\end{equation}
whose eigenvalues $\pm 1$ indicate the even/odd FP in the effective model.
Note that, since the effective Hamiltonian~\eqref{eq:effective_Hamiltonian} is constructed by extracting partial degrees of freedom from the original Hamiltonian~\eqref{eq:Hamiltonian}, the FP in the former model is in general different from that in the latter model.
For this reason, we call the eigenvalue of $(-1)^{F_{\text{eff}}}$ a \textit{partial} FP (pFP) to distinguish it from the \textit{total} FP of the full Hamiltonian~\eqref{eq:Hamiltonian}.

\subsection{Eigenvalues and Josephson current}
\begin{figure}[tbp]
 \centering
 \includegraphics[width=\linewidth]{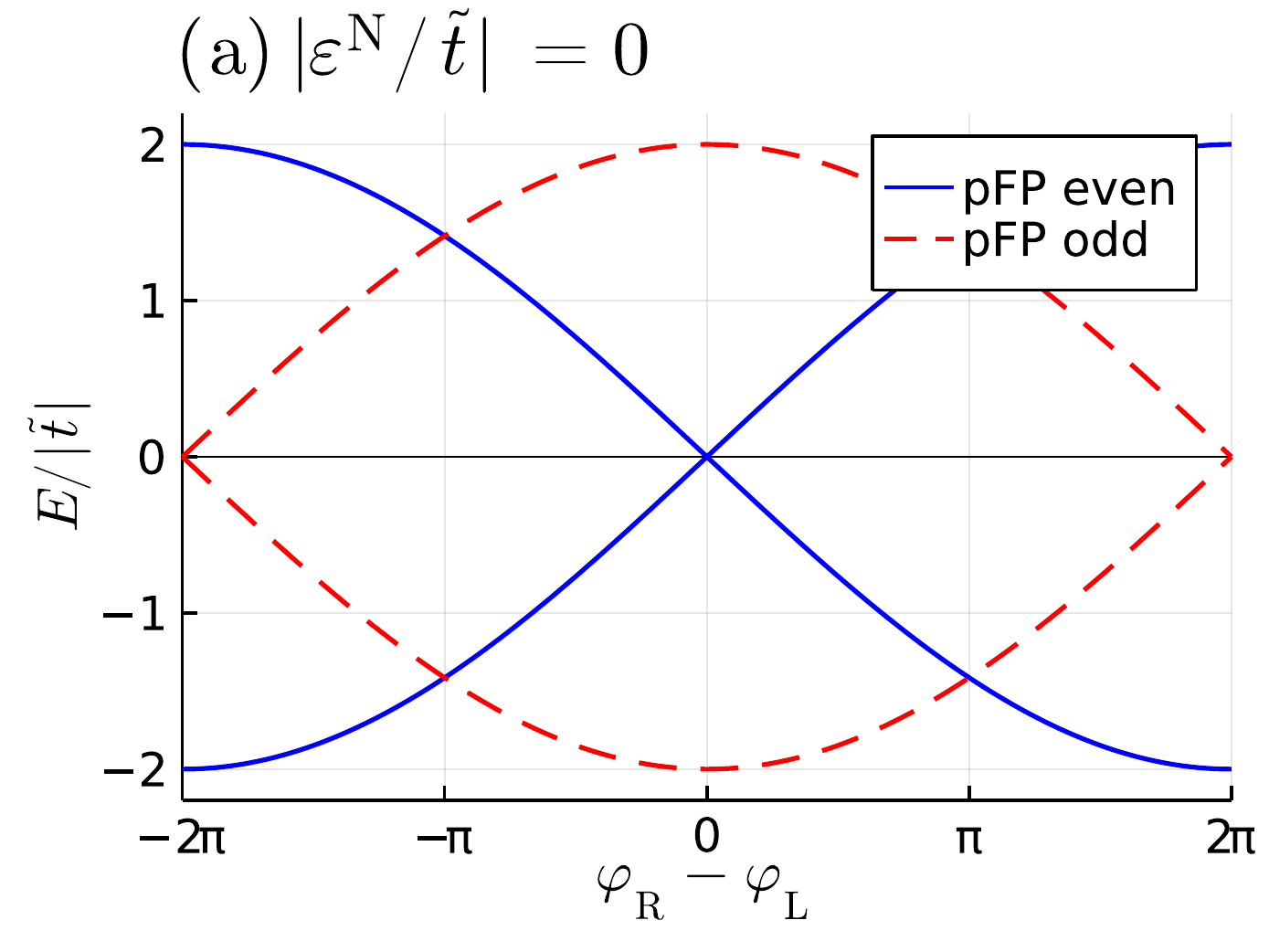}
 \includegraphics[width=\linewidth]{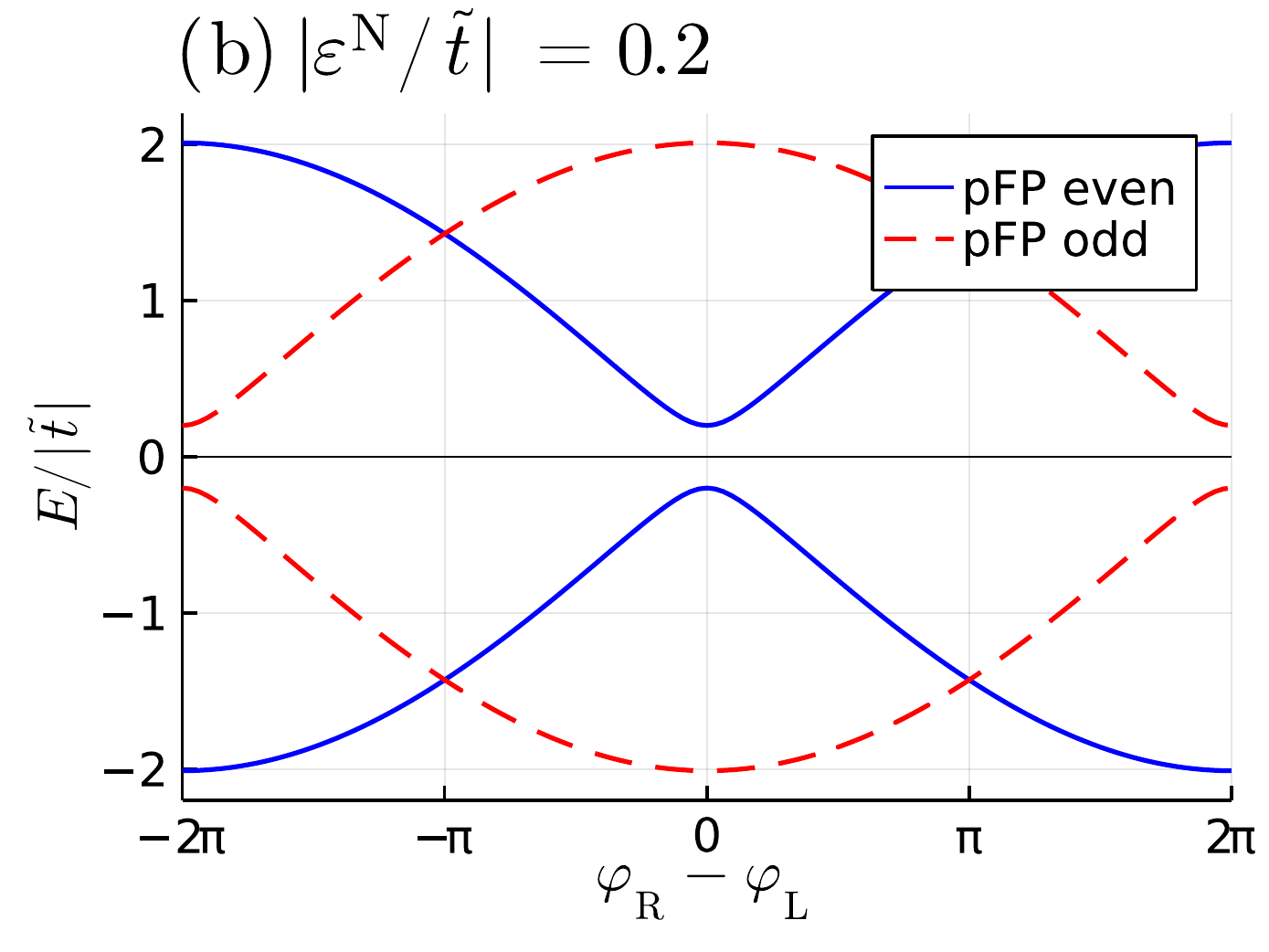}
 \includegraphics[width=\linewidth]{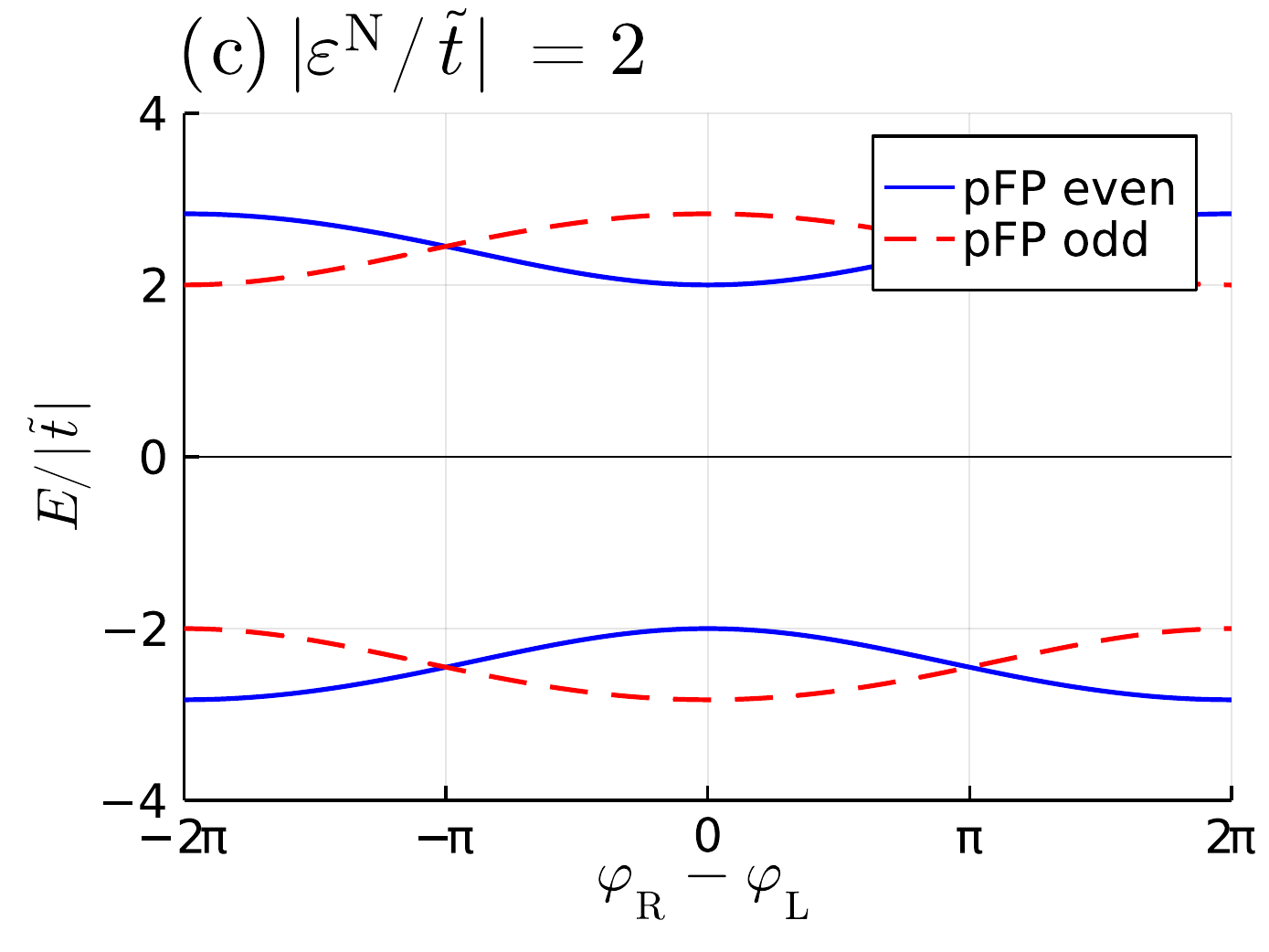}
 \caption{The $\varphi_{\text{R}} - \varphi_{\text{L}}$ dependence of energy eigenvalues in Eqs.~\eqref{eq:effective_energy_even} and \eqref{eq:effective_energy_odd} for (a) $|\varepsilon^{\text{N}} / \tilde{t}| = 0$, (b) $|\varepsilon^{\text{N}} / \tilde{t}| = 0.2$, and (c) $|\varepsilon^{\text{N}} / \tilde{t}| = 2$. $q_0$ is set to be odd. The blue solid (red dashed) lines represent the eigenvalues of the pFP-even (pFP-odd) sector. The two sectors are exchanged when $q_0$ is even.}
 \label{fig:effective_model_phidep}
\end{figure}

Next, we solve the eigenvalue problem of Eq.~\eqref{eq:effective_Hamiltonian}.
Since $[H_{\text{eff}}, (-1)^{F_{\text{eff}}}] = 0$, the effective Hamiltonian can be block-diagonalized into pFP-even and pFP-odd sectors.
Let $\ket{\text{vac}}$ be a vacuum state such that $c \ket{\text{vac}} = f \ket{\text{vac}} = 0$.
Then Eq.~\eqref{eq:effective_Hamiltonian} acts on pFP-even states as
\begin{align}
 H_{\text{eff}} \ket{\text{vac}} &= - \frac{\varepsilon^{\text{N}}}{2} \ket{\text{vac}} \notag \\
 &\quad - (\tilde{t}_{\text{L}} e^{i\varphi_{\text{L}}/2} - \tilde{t}_{\text{R}} e^{i\varphi_{\text{R}}/2}) c^\dagger f^\dagger \ket{\text{vac}}, \\
 H_{\text{eff}} c^\dagger f^\dagger \ket{\text{vac}} &= \frac{\varepsilon^{\text{N}}}{2} c^\dagger f^\dagger \ket{\text{vac}} \notag \\
 &\quad - (\tilde{t}_{\text{L}}^* e^{-i\varphi_{\text{L}}/2} - \tilde{t}_{\text{R}}^* e^{-i\varphi_{\text{R}}/2}) \ket{\text{vac}}.
\end{align}
The pFP-even sector of the Hamiltonian is thus represented by the following matrix,
\begin{equation}
 \hat{H}_{\text{eff}}^{\text{even}} =
 \begin{bmatrix}
  - \frac{\varepsilon^{\text{N}}}{2} &
  - \tilde{t}_{\text{L}}^* e^{-i\varphi_{\text{L}}/2}
  + \tilde{t}_{\text{R}}^* e^{-i\varphi_{\text{R}}/2} \\
  - \tilde{t}_{\text{L}} e^{i\varphi_{\text{L}}/2}
  + \tilde{t}_{\text{R}} e^{i\varphi_{\text{R}}/2}
  & \frac{\varepsilon^{\text{N}}}{2}
 \end{bmatrix},
\end{equation}
which has two eigenvalues,
\begin{equation}
 E^{\text{even}}_\pm =
 \pm \sqrt{ \left(\frac{\varepsilon^{\text{N}}}{2}\right)^2 + |\tilde{t}_{\text{L}}|^2 + |\tilde{t}_{\text{R}}|^2 - \left(\tilde{t}_{\text{L}} \tilde{t}_{\text{R}}^* e^{\frac{i}{2}(\varphi_{\text{L}} - \varphi_{\text{R}})} + \text{H.c.}\right) }.
 \label{eq:effective_energy_even}
\end{equation}
In a similar way, eigenvalues of the pFP-odd sector are derived as
\begin{equation}
 E^{\text{odd}}_\pm =
 \pm \sqrt{ \left(\frac{\varepsilon^{\text{N}}}{2}\right)^2 + |\tilde{t}_{\text{L}}|^2 + |\tilde{t}_{\text{R}}|^2 + \left(\tilde{t}_{\text{L}} \tilde{t}_{\text{R}}^* e^{\frac{i}{2}(\varphi_{\text{L}} - \varphi_{\text{R}})} + \text{H.c.}\right) }.
 \label{eq:effective_energy_odd}
\end{equation}
Note that both $E_\pm^{\text{even}}$ and $E_\pm^{\text{odd}}$ are $4\pi$-periodic functions of $\varphi_{\text{R}} - \varphi_{\text{L}}$.
Figure~\ref{fig:effective_model_phidep} shows the many-body energy spectrum in Eqs.~\eqref{eq:effective_energy_even} and \eqref{eq:effective_energy_odd} as functions of $\varphi_{\text{R}} - \varphi_{\text{L}}$, where we set $\tilde{t}_{\text{L}} = \tilde{t}_{\text{R}} = \tilde{t}(q_0)$ with $q_0$ chosen to be an odd integer.

Now we take a closer look at the energy spectrum with Eq.~\eqref{eq:effective_tL_tR} for different $|\varepsilon^{\text{N}} / \tilde{t}|$ values.
When $\varepsilon^{\text{N}} = 0$ [Fig.~\ref{fig:effective_model_phidep}(a)], $E_-^{\text{even/odd}}$ is reduced to the simple form
\begin{subequations}
 \label{eq:effective_model_GS_on-resonance}
 \begin{align}
  E_{-}^{\text{even}} &= -2 |\tilde{t}(q_0)| \left|\sin\left(\frac{\varphi_{\text{R}} - \varphi_{\text{L}}}{4}\right) \right|, \\
  E_{-}^{\text{odd}} &= -2 |\tilde{t}(q_0)| \left|\cos\left(\frac{\varphi_{\text{R}} - \varphi_{\text{L}}}{4}\right) \right|,
 \end{align}
\end{subequations}
where $q_0$ is assumed to be an odd integer, i.e., $\tilde{t}_{\text{L}} = \tilde{t}_{\text{R}}$.
When $q_0$ is an even integer ($\tilde{t}_{\text{L}} = -\tilde{t}_{\text{R}}$), $E_-^{\text{even}}$ and $E_-^{\text{odd}}$ are exchanged.
Equation~\eqref{eq:effective_model_GS_on-resonance} is consistent with the result of the first-order perturbation theory [Eq.~\eqref{eq:1st_perturbed_energy_GS}].
Note that the lowest energy in each pFP sector $E_-^{\text{even/odd}}$ is linear in the tunneling coupling $\lambda t$, i.e., the interaction energy between the two MZMs is proportional to the tunneling amplitude of a single fermion, rather than to tunneling probability of a single fermion or to tunneling amplitude of a Cooper pair; the latter two are proportional to $(\lambda t)^2$.
At finite temperature the free energy of $H_{\text{eff}}$ in each pFP sector is given by
\begin{subequations}
 \begin{align}
  F^{\text{even}} &= -\frac{1}{\beta} \ln\left\{2\cosh\left[2\beta |\tilde{t}(q_0)| \sin\left(\frac{\varphi_{\text{R}} - \varphi_{\text{L}}}{4}\right)\right]\right\}, \\
  F^{\text{odd}} &=  -\frac{1}{\beta} \ln\left\{2\cosh\left[2\beta |\tilde{t}(q_0)| \cos\left(\frac{\varphi_{\text{R}} - \varphi_{\text{L}}}{4}\right)\right]\right\},
 \end{align}
\end{subequations}
where $\beta$ is inverse temperature.
Obviously, the free energy has a period $4\pi$ in the phase difference $\varphi_{\text{R}} - \varphi_{\text{L}}$.
The dc Josephson current is obtained by differentiating the free energy $F$ with respect to the phase difference,
\begin{equation}
  I = \frac{e}{\hbar}\frac{\partial F}{\partial(\varphi_{\text{R}} - \varphi_{\text{L}})}.
\end{equation}
In particular, in the zero-temperature limit $\beta\to\infty$, the dc Josephson current is
\begin{subequations}
 \label{eq:eff_Josephson_current_on-resonance}
 \begin{align}
  I^{\text{even}} &= -\frac{e|\tilde{t}(q_0)|}{\hbar}
  \sgn\left[\sin\left(\frac{\varphi_{\text{R}} - \varphi_{\text{L}}}{4}\right)\right]
  \cos\left(\frac{\varphi_{\text{R}} - \varphi_{\text{L}}}{4}\right), \\
  I^{\text{odd}} &= +\frac{e|\tilde{t}(q_0)|}{\hbar}
  \sgn\left[\cos\left(\frac{\varphi_{\text{R}} - \varphi_{\text{L}}}{4}\right)\right]
  \sin\left(\frac{\varphi_{\text{R}} - \varphi_{\text{L}}}{4}\right).
  \end{align}
\end{subequations}
The Josephson current is a $4\pi$-periodic function, which becomes discontinuous in the limit $\beta \to \infty$ at $\varphi_{\text{R}} - \varphi_{\text{L}} = 4n\pi$ in $I^{\text{even}}$ and at $\varphi_{\text{R}} - \varphi_{\text{L}} = (4n+2)\pi$ in $I^{\text{odd}}$ ($n \in \mathbb{Z}$).
The discontinuities are the consequence of the level crossings at $\varphi_{\text{R}} - \varphi_{\text{L}} = 2n\pi$ in Fig.~\ref{fig:effective_model_phidep}(a).

These level crossings are lifted when $\varepsilon^{\text{N}} \neq 0$ [see Fig.~\ref{fig:effective_model_phidep}(b)].
In other words, the gap opens at $E = 0$ when the single-particle energy levels in the middle N region are not in resonance with the MZMs.
When $|\varepsilon^{\text{N}}|$ is much larger than $|\tilde{t}(q_0)|$ [Fig.~\ref{fig:effective_model_phidep}(c)], the $\varphi_{\text{R}} - \varphi_{\text{L}}$ dependence of the lowest energy eigenvalue in each pFP sector is given by
\begin{subequations}
 \label{eq:effective_model_GS_off-resonance}
 \begin{align}
  E_-^{\text{even}} &\simeq -\frac{|\varepsilon^{\text{N}}|}{2}
  - \frac{2|\tilde{t}(q_0)|^2}{|\varepsilon^{\text{N}}|}
  \left[1 + (-1)^{q_0} \cos\left(\frac{\varphi_{\text{R}} - \varphi_{\text{L}}}{2}\right)\right],
  \\
  E_-^{\text{odd}} &\simeq -\frac{|\varepsilon^{\text{N}}|}{2}
  - \frac{2|\tilde{t}(q_0)|^2}{|\varepsilon^{\text{N}}|}
  \left[1 - (-1)^{q_0} \cos\left(\frac{\varphi_{\text{R}} - \varphi_{\text{L}}}{2}\right)\right],
 \end{align}
\end{subequations}
which is consistent with the result of the second-order perturbation theory [Eq.~\eqref{eq:2nd_perturbed_energy_GS}].
The dc Josephson current at zero temperature is thus given by
\begin{equation}
 I = (-1)^{F_{\text{eff}} + q_0} \frac{e |\tilde{t}(q_0)|^2}{\hbar|\varepsilon^{\text{N}}|} \sin\left(\frac{\varphi_{\text{R}} - \varphi_{\text{L}}}{2}\right),
 \label{eq:eff_Josephson_current_off-resonance}
\end{equation}
As a result, we have the $4\pi$-periodic Josephson effect~\cite{Kitaev2001, Kwon2004, Fu2009, Lutchyn2010, Nogueira2012, Zhang2014_PRB, Affleck2014} in the on-resonance case as well as the off-resonance case.
Of course, the magnitude of the Josephson current in the on-resonance case [Eq.~\eqref{eq:eff_Josephson_current_on-resonance}] is much larger than that in the off-resonance case [Eq.~\eqref{eq:eff_Josephson_current_off-resonance}].

\subsection{Discussions about fermion parity}
\label{sec:effective_model_FP}
In the previous subsection, we have solved the effective model connecting the on- and off-resonance cases, and derived the explicit form of the Josephson current for both cases [Eqs.~\eqref{eq:eff_Josephson_current_on-resonance} and \eqref{eq:eff_Josephson_current_off-resonance}].
Comparing these two equations, we notice that Eq.~\eqref{eq:eff_Josephson_current_off-resonance} has a $(-1)^{q_0}$ factor whereas Eq.~\eqref{eq:eff_Josephson_current_on-resonance} does not.
One might consider that the $(-1)^{q_0}$ factor would result in alternating behavior of the Josephson current (for some fixed phase difference) as a function of the chemical potential $\mu$, with sign changes happening near resonance points.
However, this would not match enhanced Josephson current in the on-resonance case \eqref{eq:eff_Josephson_current_on-resonance}.
This potential inconsistency is resolved by considering the FP, as we explain below.
Recall that the pFP is different from the total FP of the original Hamiltonian~\eqref{eq:Hamiltonian}.
When the single-particle states with $q = 1, \dots, q_0-1$ in the N region are occupied and those with $q > q_0$ are empty, the total FP $(-1)^F$ is related to the pFP $(-1)^{F_{\text{eff}}}$ by
\begin{equation}
 (-1)^F = (-1)^{F_{\text{eff}}} (-1)^{q_0-1}.
 \label{eq:FP-pFP_relation}
\end{equation}
Here we note that the pFP $F_{\text{eff}}$ is the FP of the $q = q_0$ state and the MZMs [Eq.~\eqref{eq:partial_FP}].
Changing the chemical potential can change the FP of all the single-particle states in the N region, which is compensated by the FP of the MZMs
\begin{equation}
 1 - 2f^\dagger f = i\gamma_{\text{R}} \gamma_{\text{L}},
 \label{eq:FP_MZMs}
\end{equation}
so that the total FP is conserved.

In the off-resonance case where $|\varepsilon^{\text{N}}| \gg |\tilde{t}_{\text{L/R}}|$, we can integrate out the normal level with energy $\varepsilon^{\text{N}}$ in the effective Hamiltonian $H_{\text{eff}}$ and obtain the effective Hamiltonian for the MZMs
\begin{align}
 H_{\text{eff}}^{\text{off}} &=
 - \frac{1}{|\varepsilon^{\text{N}}|}(|\tilde{t}_{\text{L}}|^2 + |\tilde{t}_{\text{R}}|^2) \notag \\
 & \quad -i \gamma_{\text{R}} \gamma_{\text{L}} \frac{2(-1)^{q_0} |\tilde{t}_{\text{L}}\tilde{t}_{\text{R}}|}{\varepsilon^{\text{N}}}
 \cos\!\left(\frac{\varphi_{\text{R}} - \varphi_{\text{L}}}{2}\right),
\end{align}
which corresponds to the $1/|\varepsilon^{\text{N}}|$ terms in Eq.~\eqref{eq:effective_model_GS_off-resonance}.
We can generalize it to include the contribution from all the normal modes ($q = 1, \dots, L$),
\begin{equation}
 H^{\text{off}} =
 i \gamma_{\text{R}} \gamma_{\text{L}}
 \lambda^2 t A^{(2)}(\bar{\mu}, \bar{\Delta}, L)
 \cos\!\left(\frac{\varphi_{\text{R}} - \varphi_{\text{L}}}{2}\right),
 \label{eq:effective_Hamiltonian_off-resonance}
\end{equation}
where $A^{(2)}(\bar{\mu}, \bar{\Delta}, L)$ is defined in Eq.~\eqref{eq:2nd_perturbed_amplitude} and we have kept only the terms related to the FP of the MZMs.
Differentiating \eqref{eq:effective_Hamiltonian_off-resonance} with respect to the phase difference gives the Josephson current,
\begin{equation}
 I^{\text{off}} = -i \gamma_{\text{R}} \gamma_{\text{L}}
 \frac{e\lambda^2 t}{\hbar} A^{(2)}(\bar{\mu}, \bar{\Delta}, L)
 \sin\!\left(\frac{\varphi_{\text{R}} - \varphi_{\text{L}}}{2}\right).
\label{eq:Josephson_current_off-resonance}
\end{equation}
As we have seen in Fig.~\ref{fig:2nd_perturbed_amplitude}, the function $A^{(2)}(\bar{\mu}, \bar{\Delta}, L)$ has the alternating dependence on $\mu$, and its sign is equal to the FP of the single-particle states in the N region $(-1)^{F^{\text{N}}}$.
Since the product of $i\gamma_{\text{R}} \gamma_{\text{L}}$ and $(-1)^{F^{\text{N}}}$ is the total FP $(-1)^F$, Eq.~\eqref{eq:Josephson_current_off-resonance} can be rewritten as
\begin{equation}
 I^{\text{off}} = (-1)^{1+F}
 \frac{e\lambda^2 t}{\hbar} \left|A^{(2)}(\bar{\mu}, \bar{\Delta}, L)\right|
 \sin\!\left(\frac{\varphi_{\text{R}} - \varphi_{\text{L}}}{2}\right).
 \label{eq:Josephson_current_off-resonance 2}
\end{equation}
Equation~\eqref{eq:Josephson_current_off-resonance 2} is a generalization of Eq.~\eqref{eq:eff_Josephson_current_off-resonance} for the off-resonance case.
Note that the Josephson current in Eq.~\eqref{eq:Josephson_current_off-resonance 2} is proportional to the total FP $(-1)^F$ and does not oscillate with the chemical potential $\mu$ as long as the total FP is conserved.

Remembering the relation~\eqref{eq:FP-pFP_relation} and that Eq.~\eqref{eq:eff_Josephson_current_on-resonance} are written for odd $q_0$, we can restate Eq.~\eqref{eq:eff_Josephson_current_on-resonance} in terms of the total FP; the dc Josephson current through a resonant level $q_0$ is given by
\begin{subequations}
 \label{eq:Josephson_current_on-resonance}
 \begin{equation}
  I^{\text{on}} =
  - \frac{e|\tilde{t}(q_0)|}{\hbar}
   \sgn\!\left[\sin\left(\frac{\varphi_{\text{R}} - \varphi_{\text{L}}}{4}\right)\right]
   \cos\left(\frac{\varphi_{\text{R}} - \varphi_{\text{L}}}{4}\right)
 \end{equation} 
 for $(-1)^F = +1$, and
 \begin{equation}
 I^{\text{on}} =
  +\frac{e|\tilde{t}(q_0)|}{\hbar}
   \sgn\!\left[\cos\left(\frac{\varphi_{\text{R}} - \varphi_{\text{L}}}{4}\right)\right]
   \sin\left(\frac{\varphi_{\text{R}} - \varphi_{\text{L}}}{4}\right)
 \end{equation}
\end{subequations}
for $(-1)^F = -1$.
We note that Eqs.~\eqref{eq:Josephson_current_off-resonance 2} and \eqref{eq:Josephson_current_on-resonance} are consistent with each other in that they both have the same sign as $(-1)^{1+F}$ for $0 < \varphi_{\text{R}} - \varphi_{\text{L}} \ll \pi$.

We comment on the related previous work of Affleck and Giuliano~\cite{Affleck2014}.
Assuming the total-FP conservation~%
\footnote{In Ref.~\cite{Affleck2014}, they have also considered the case where the total FP is not conserved, obtaining $2\pi$-periodic Josephson current.},
these authors have derived the current--phase relation equivalent to Eqs.~\eqref{eq:Josephson_current_off-resonance 2} and \eqref{eq:Josephson_current_on-resonance} in the short-junction limit where only a \textit{single} level in the N region is involved in the supercurrent flow.
Here we have calculated the Josephson current in the presence of \textit{several} occupied levels in the N region and obtained the dependence on the total FP $(-1)^F$, while taking into account the oscillatory $(-1)^{q_0 - 1}$ factor of the wavefunctions in the N region of finite length.

We have obtained the expressions for the dc Josephson current for the on- and off-resonance cases for the SNS Josephson junction with multiple levels in the N region.
In the next section we will develop a numerical method that allows us to quantitatively describe how these limiting formulas are smoothly connected.

\section{Exact diagonalization of corner-modified banded block-Toeplitz matrix}
\label{sec:exact_diagonalization}
In Secs.~\ref{sec:perturbation} and \ref{sec:effective_model}, we have investigated the interaction between the two MZMs and the Josephson current using the perturbation theory.
In this section we introduce a nonperturbative, numerical exact diagonalization method for a more quantitative analysis of the energy structures.
Indeed, the Hamiltonian~\eqref{eq:Hamiltonian} has the form of a \textit{corner-modified banded block-Toeplitz matrix}, for which an efficient numerical diagonalization method has been developed in recent theoretical studies~\cite{Alase2016, Cobanera2017, Alase2017, Cobanera2018}.
We thus apply the method to our model, and compare its numerical results with the perturbation theory.

\subsection{Preliminary}
Let us consider a problem equivalent to the exact diagonalization of the Hamiltonian~\eqref{eq:Hamiltonian}, in the framework of the corner-modified banded block-Toeplitz matrix~\cite{Alase2016, Cobanera2017, Alase2017, Cobanera2018}.
In the method, we first fix a certain initial energy value $\varepsilon$, and then construct wavefunctions preserving the translation symmetry, which are called bulk solutions, in each region of SL, SR, and N.
Next, we obtain a boundary matrix $B(\varepsilon)$ taking into account boundary (interface) conditions between the bulk solutions~%
\footnote{In terms of the corner-modified banded block-Toeplitz matrix, boundary conditions correspond to a \textit{corner-modification} of the matrix.}.
When $\det B(\varepsilon) = 0$, $\varepsilon$ is an eigenenergy of the original Hamiltonian, which is exactly what we want.
Otherwise, it is possible to make $\varepsilon$ converge to a value satisfying $\det B(\varepsilon) = 0$, by using some standard numerical technique.
Therefore, the method can be systematically implemented in numerical calculations, whose results are shown in Sec.~\ref{sec:exact_diagonalization_numerics}.

\subsubsection{Bulk solutions}
Here, bulk solutions of the model are constructed as follows.
Let us consider a certain value $\varepsilon \in \mathbb{R}$ as a given energy parameter.
For the calculations of the bulk solutions, it is convenient to treat \textit{translation-invariant auxiliary Hamiltonians}~\cite{Alase2017, Cobanera2017},
\begin{align}
 \bm{H}_{\text{S}}(\varphi) &= \bm{1} \otimes h_0^{\text{S}} + \left( \bm{T} \otimes h_1^{\text{S}}(\varphi) + \bm{T}^{-1} \otimes h_1^{\text{S}}(\varphi)^\dagger \right),
 \label{eq:auxiliary_Hamiltonian_S} \\
 \bm{H}_{\text{N}} &= \bm{1} \otimes h_0^{\text{N}} + \left( \bm{T} \otimes h_1^{\text{N}} + \bm{T}^{-1} \otimes h_1^{\text{N} \dagger} \right),
 \label{eq:auxiliary_Hamiltonian_N}
\end{align}
rather than the original Hamiltonian [Eqs.~\eqref{eq:Hamiltonian} and \eqref{eq:Hamiltonian_SL_1st}--\eqref{eq:Hamiltonian_T_1st}].
Here $\bm{T} = \sum_{j\in\mathbb{Z}} \ketbra{j}{j+1}$ denotes a generator of discrete translations, and $\bm{1} = \sum_{j\in\mathbb{Z}} \ketbra{j}{j}$ is a corresponding identity operator.
More specifically, we solve the following eigenvalue equations,
\begin{align}
 \bm{H}_{\text{S}}(\varphi) \Ket{\Psi^{\text{S}}(\varepsilon; \varphi)} &= \varepsilon \Ket{\Psi^{\text{S}}(\varepsilon; \varphi)}, \\
 \bm{H}_{\text{N}} \Ket{\Psi^{\text{N}}(\varepsilon)} &= \varepsilon \Ket{\Psi^{\text{N}}(\varepsilon)},
\end{align}
where the eigenvectors $\Ket{\Psi^{\text{S}}(\varepsilon; \varphi)}$ and $\Ket{\Psi^{\text{N}}(\varepsilon)}$ possess ``translation symmetry'', since $\bm{H}_{\text{S}}(\varphi)$ and $\bm{H}_{\text{N}}$ represent the Hamiltonian of an infinite system without boundaries (interfaces).
Then the bulk solutions are defined by~%
\footnote{Strictly speaking, it is necessary to restrict the bulk solutions so that $\Braket{\psi^{\text{SL}}(\varepsilon; \varphi_{\text{L}}) | \psi^{\text{SL}}(\varepsilon; \varphi_{\text{L}})}, \Braket{\psi^{\text{SR}}(\varepsilon; \varphi_{\text{R}}) | \psi^{\text{SR}}(\varepsilon; \varphi_{\text{R}})} < \infty$, since the SL and SR regions are semi-infinite systems.}
\begin{align}
 \Ket{\psi^{\text{SL}}(\varepsilon; \varphi_{\text{L}})} &:= \left( \sum_{j=-\infty}^{0} \ketbra{j}{j} \otimes \mathbbm{1}_2 \right) \Ket{\Psi^{\text{S}}(\varepsilon; \varphi_{\text{L}})}, \\
 \Ket{\psi^{\text{SR}}(\varepsilon; \varphi_{\text{R}})} &:= \left( \sum_{j=L+1}^{\infty} \ketbra{j}{j} \otimes \mathbbm{1}_2 \right) \Ket{\Psi^{\text{S}}(\varepsilon; \varphi_{\text{R}})}, \displaybreak[2] \\
 \Ket{\psi^{\text{N}}(\varepsilon)} &:= \left( \sum_{j=1}^{L} \ketbra{j}{j} \otimes \mathbbm{1}_2 \right) \Ket{\Psi^{\text{N}}(\varepsilon)}.
\end{align}

Now we show concrete expressions of the wavefunctions.
In the left and right superconducting regions, the bulk solutions for an energy $\varepsilon$ are given by
\begin{align}
 \Ket{\psi^{\text{SL}}_m(\varepsilon; \varphi_{\text{L}})} &= \sum_{j=-\infty}^{0} \left[z^{\text{S}}_m(\varepsilon)\right]^{-j} \ket{j} \Ket{u^{\text{S}}_m(\varepsilon; \varphi_{\text{L}})}, \displaybreak[2] \\
 \Ket{\psi^{\text{SR}}_m(\varepsilon; \varphi_{\text{R}})} &= \sum_{j=L+1}^{\infty} \left[z^{\text{S}}_m(\varepsilon)\right]^{j-(L+1)} \ket{j} \Ket{u^{\text{S}}_m(\varepsilon; \varphi_{\text{R}})},
\end{align}
with $m = 1, 2$.
For the definition of $z^{\text{S}}_m(\varepsilon)$ and $\Ket{u^{\text{S}}_m(\varepsilon; \varphi)}$, see Eqs.~\eqref{eq:z_sc}--\eqref{eq:z_sc_inequality} in Appendix~\ref{app:exact_sol}.
In the N region, on the other hand, the bulk solutions are given by
\begin{equation}
 \Ket{\psi^{\text{N}}_{l \sigma}(\varepsilon)} = \sum_{j=1}^{L} \left[z^{\text{N}}_l(\varepsilon)\right]^{\sigma j} \ket{j} \Ket{u^{\text{N}}_l},
\end{equation}
where $l = 1, 2$ and $\sigma = \pm$.
$z^{\text{N}}_l(\varepsilon)$ and $\Ket{u^{\text{N}}_l}$ are defined by Eqs~\eqref{eq:z_n} and \eqref{eq:u_n}, respectively.

\subsubsection{Boundary matrix}
In the above construction of the bulk solutions, the regions SL, SR, and N have been independently considered.
The tunneling Hamiltonian $H_{\text{T}}$, however, has matrix elements connecting the wavefunctions in the neighboring regions.
Therefore, we here construct a boundary matrix reflecting the boundary (interface) conditions.
Let $H_\varepsilon$ be $H - \varepsilon \bm{1}$.
Then the $8 \times 8$ boundary matrix~%
\footnote{Remark: the wavefunctions ($\Ket{\psi^{\text{SL}}_m(\varepsilon; \varphi_{\text{L}})}$, $\Ket{\psi^{\text{SR}}_m(\varepsilon; \varphi_{\text{R}})}$, and $\Ket{\psi^{\text{N}}_m(\varepsilon)}$) have two internal degrees of freedom (particle and hole).}
is defined by
\begin{equation}
 B(\varepsilon) =
 \begin{bmatrix}
  B^{\text{SL}}(\varepsilon) & B^{\text{N}}(\varepsilon) & B^{\text{SR}}(\varepsilon)
 \end{bmatrix},
 \label{eq:boundary_matrix}
\end{equation}
where
\begin{widetext}
 \begin{subequations}
  \begin{align}
   B^{\text{SL}}(\varepsilon) &=
   \begin{bmatrix}
    \Braket{0 | H_\varepsilon | \psi^{\text{SL}}_1(\varepsilon; \varphi_{\text{L}})} & \Braket{0 | H_\varepsilon | \psi^{\text{SL}}_2(\varepsilon; \varphi_{\text{L}})} \\[2mm]
    \Braket{1 | H_\varepsilon | \psi^{\text{SL}}_1(\varepsilon; \varphi_{\text{L}})} & \Braket{1 | H_\varepsilon | \psi^{\text{SL}}_2(\varepsilon; \varphi_{\text{L}})} \\[2mm]
    \Braket{L | H_\varepsilon | \psi^{\text{SL}}_1(\varepsilon; \varphi_{\text{L}})} & \Braket{L | H_\varepsilon | \psi^{\text{SL}}_2(\varepsilon; \varphi_{\text{L}})} \\[2mm]
    \Braket{L+1 | H_\varepsilon | \psi^{\text{SL}}_1(\varepsilon; \varphi_{\text{L}})} & \Braket{L+1 | H_\varepsilon | \psi^{\text{SL}}_2(\varepsilon; \varphi_{\text{L}})}
   \end{bmatrix}, \displaybreak[2]
   \label{eq:boundary_matrix_SL} \\
   B^{\text{N}}(\varepsilon) &=
   \begin{bmatrix}
    \Braket{0 | H_\varepsilon | \psi^{\text{N}}_{1+}(\varepsilon)} & \Braket{0 | H_\varepsilon | \psi^{\text{N}}_{1-}(\varepsilon)} & \Braket{0 | H_\varepsilon | \psi^{\text{N}}_{2+}(\varepsilon)} & \Braket{0 | H_\varepsilon | \psi^{\text{N}}_{2-}(\varepsilon)} \\[2mm]
    \Braket{1 | H_\varepsilon | \psi^{\text{N}}_{1+}(\varepsilon)} & \Braket{1 | H_\varepsilon | \psi^{\text{N}}_{1-}(\varepsilon)} & \Braket{1 | H_\varepsilon | \psi^{\text{N}}_{2+}(\varepsilon)} & \Braket{1 | H_\varepsilon | \psi^{\text{N}}_{2-}(\varepsilon)} \\[2mm]
    \Braket{L | H_\varepsilon | \psi^{\text{N}}_{1+}(\varepsilon)} & \Braket{L | H_\varepsilon | \psi^{\text{N}}_{1-}(\varepsilon)} & \Braket{L | H_\varepsilon | \psi^{\text{N}}_{2+}(\varepsilon)} & \Braket{L | H_\varepsilon | \psi^{\text{N}}_{2-}(\varepsilon)} \\[2mm]
    \Braket{L+1 | H_\varepsilon | \psi^{\text{N}}_{1+}(\varepsilon)} & \Braket{L+1 | H_\varepsilon | \psi^{\text{N}}_{1-}(\varepsilon)} & \Braket{L+1 | H_\varepsilon | \psi^{\text{N}}_{2+}(\varepsilon)} & \Braket{L+1 | H_\varepsilon | \psi^{\text{N}}_{2-}(\varepsilon)}
   \end{bmatrix}, \displaybreak[2]
   \label{eq:boundary_matrix_N} \\
   B^{\text{SR}}(\varepsilon) &=
   \begin{bmatrix}
    \Braket{0 | H_\varepsilon | \psi^{\text{SR}}_1(\varepsilon; \varphi_{\text{R}})} & \Braket{0 | H_\varepsilon | \psi^{\text{SR}}_2(\varepsilon; \varphi_{\text{R}})} \\[2mm]
    \Braket{1 | H_\varepsilon | \psi^{\text{SR}}_1(\varepsilon; \varphi_{\text{R}})} & \Braket{1 | H_\varepsilon | \psi^{\text{SR}}_2(\varepsilon; \varphi_{\text{R}})} \\[2mm]
    \Braket{L | H_\varepsilon | \psi^{\text{SR}}_1(\varepsilon; \varphi_{\text{R}})} & \Braket{L | H_\varepsilon | \psi^{\text{SR}}_2(\varepsilon; \varphi_{\text{R}})} \\[2mm]
    \Braket{L+1 | H_\varepsilon | \psi^{\text{SR}}_1(\varepsilon; \varphi_{\text{R}})} & \Braket{L+1 | H_\varepsilon | \psi^{\text{SR}}_2(\varepsilon; \varphi_{\text{R}})}
   \end{bmatrix}.
   \label{eq:boundary_matrix_SR}
  \end{align}
 \end{subequations}
\end{widetext}
The matrix elements of Eqs.~\eqref{eq:boundary_matrix_SL}--\eqref{eq:boundary_matrix_SR} are calculated as
\begin{subequations}
 \begin{align}
  \Braket{0 | H_\varepsilon | \psi^{\text{SL}}_m(\varepsilon; \varphi_{\text{L}})} &= - z^{\text{S}}_m(\varepsilon) h^{\text{S}}_1(\varphi_{\text{L}}) \Ket{u^{\text{S}}_m(\varepsilon; \varphi_{\text{L}})}, \\
  \Braket{1 | H_\varepsilon | \psi^{\text{SL}}_m(\varepsilon; \varphi_{\text{L}})} &= \lambda h^{\text{N} \dagger}_1 \Ket{u^{\text{S}}_m(\varepsilon; \varphi_{\text{L}})}, \displaybreak[2] \\
  \Braket{0 | H_\varepsilon | \psi^{\text{N}}_{l \sigma}(\varepsilon)} &= \left[z^{\text{N}}_l(\varepsilon)\right]^\sigma \lambda h^{\text{N}}_1 \Ket{u^{\text{N}}_l}, \\
  \Braket{1 | H_\varepsilon | \psi^{\text{N}}_{l \sigma}(\varepsilon)} &= - h^{\text{N} \dagger}_1 \Ket{u^{\text{N}}_l}, \\
  \Braket{L | H_\varepsilon | \psi^{\text{N}}_{l \sigma}(\varepsilon)} &= - \left[z^{\text{N}}_l(\varepsilon)\right]^{\sigma (L+1)} h^{\text{N}}_1 \Ket{u^{\text{N}}_l}, \\
  \Braket{L+1 | H_\varepsilon | \psi^{\text{N}}_{l \sigma}(\varepsilon)} &= \left[z^{\text{N}}_l(\varepsilon)\right]^{\sigma L} \lambda h^{\text{N} \dagger}_1 \Ket{u^{\text{N}}_l}, \displaybreak[2] \\
  \Braket{L | H_\varepsilon | \psi^{\text{SR}}_m(\varepsilon; \varphi_{\text{R}})} &= \lambda h^{\text{N}}_1 \Ket{u^{\text{S}}_m(\varepsilon; \varphi_{\text{R}})}, \\
  \Braket{L+1 | H_\varepsilon | \psi^{\text{SR}}_m(\varepsilon; \varphi_{\text{R}})} &= - \left[z^{\text{S}}_m(\varepsilon)\right]^{-1} h^{\text{S}}_1(\varphi_{\text{R}})^\dagger \Ket{u^{\text{S}}_m(\varepsilon; \varphi_{\text{R}})},
 \end{align}
\end{subequations}
with $m = 1, 2$; $l = 1, 2$, and $\sigma = \pm$.
The other matrix elements are zero.

As mentioned above, if a value $\varepsilon$ satisfying $\det B(\varepsilon) = 0$ is found, then it is an exact eigenenergy of the original Hamiltonian~\eqref{eq:Hamiltonian}.
In this case there exists a nontrivial kernel $\bm{\alpha}$ of the boundary matrix [i.e., $B(\varepsilon) \bm{\alpha} = 0$], which enables us to construct a wavefunction corresponding to the exact eigenvalue~\cite{Alase2016, Cobanera2017, Alase2017, Cobanera2018}.

\subsection{Numerical results}
\label{sec:exact_diagonalization_numerics}

In this subsection, we present numerical results of exact eigenenergies of the Hamiltonian~\eqref{eq:Hamiltonian}.
The energies where the boundary matrix [Eq.~\eqref{eq:boundary_matrix}] has a nontrivial kernel can be obtained by using a conventional root-finding algorithm, such as Newton's method.
The detailed results are shown below.

Figures~\ref{fig:phidep_exact}(a) and \ref{fig:phidep_exact}(c) represent single-particle energy spectra (around $\varepsilon = 0$) of the exact eigenvalues at $\mu = 0$, as functions of $\varphi_{\text{R}} - \varphi_{\text{L}}$.
As noted in the beginning of Sec.~\ref{sec:perturbation_on-resonance}, the zero chemical potential corresponds to on-resonance (off-resonance) regime for odd (even) $L$.
Indeed, the difference between even and odd $L$s is clearly seen; for $L = 5$ [Fig.~\ref{fig:phidep_exact}(a)], the $\varphi_{\text{R}} - \varphi_{\text{L}}$ dependence is consistent with that of the first-order perturbed energy in Eq.~\eqref{eq:1st_perturbed_energy}, while the dependence for $L = 6$ [Fig.~\ref{fig:phidep_exact}(c)] is in good agreement with the second-order perturbation theory [Eq.~\eqref{eq:2nd_perturbed_energy}].
When the chemical potential slightly deviates from the on-resonance regime [Fig.~\ref{fig:phidep_exact}(b)], the level crossings at $\varphi_{\text{R}} - \varphi_{\text{L}} = 2n\pi$ are lifted, while those at $\varphi_{\text{R}} - \varphi_{\text{L}} = (2n+1)\pi$ remain because of protection by the FP.
This corresponds to the many-body energy spectrum in Fig.~\ref{fig:effective_model_phidep}(b), which has small gaps at $\varphi_{\text{R}} - \varphi_{\text{L}} = 2n\pi$ around the zero energy.
Moreover, the magnitude of the energy is on the order of $\lambda t$ and $\lambda^2 t$ for on- and off-resonance cases, respectively.
Therefore, our perturbation theory in Sec.~\ref{sec:perturbation} describes well the energy structures around the zero energy.

\begin{figure}[tbp]
 \includegraphics[width=.95\linewidth]{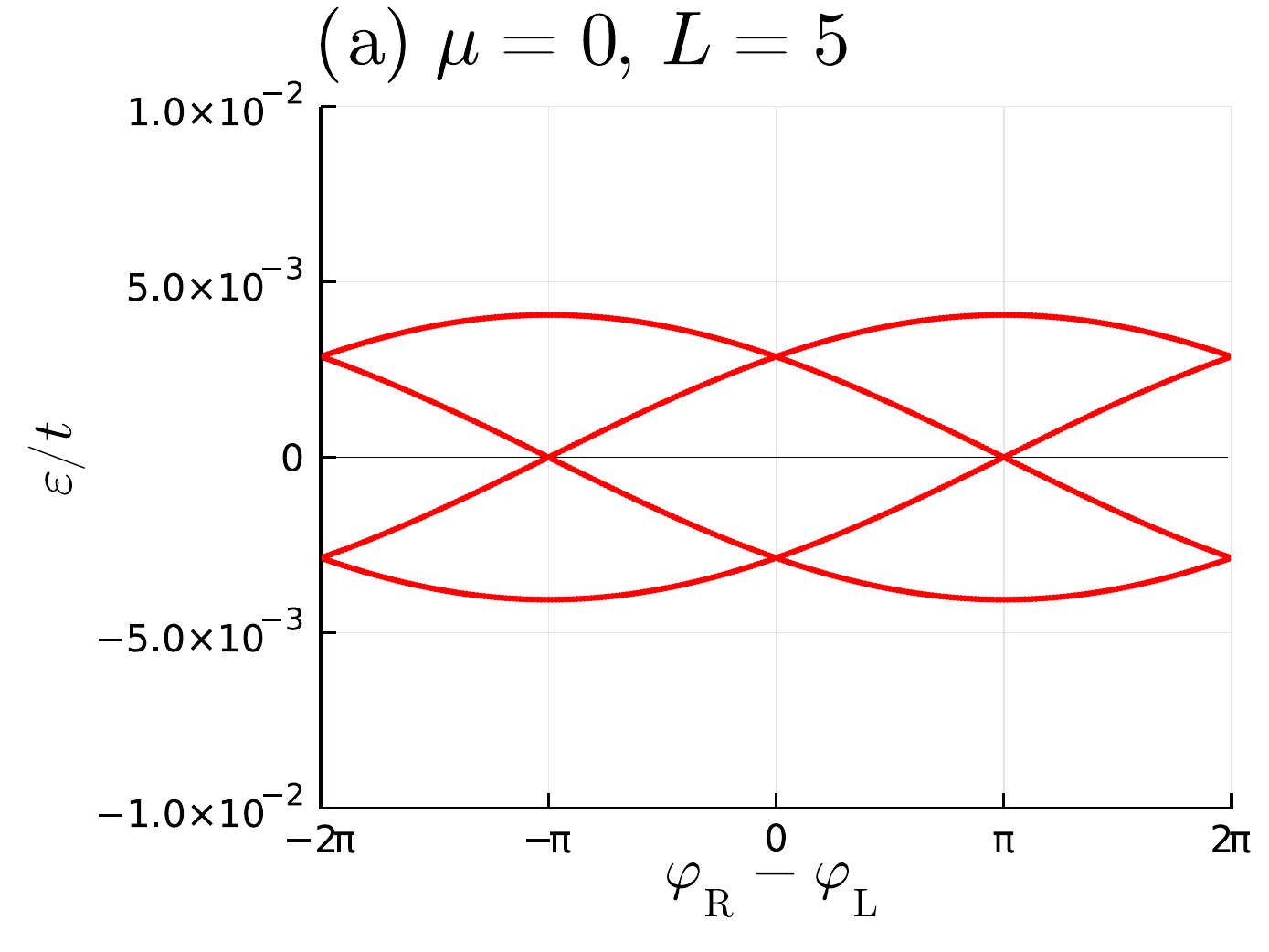}
 \includegraphics[width=.95\linewidth]{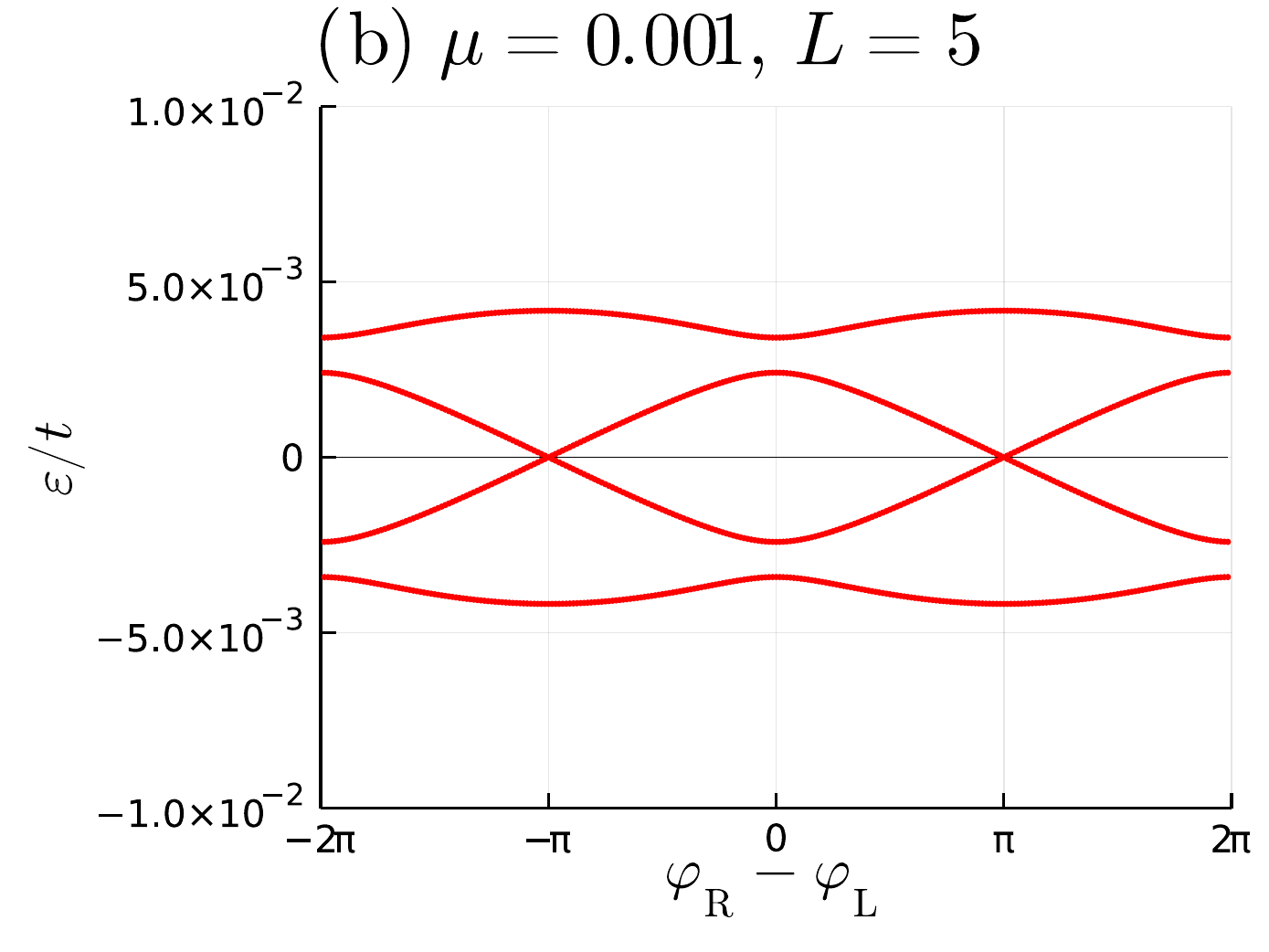}
 \includegraphics[width=.95\linewidth]{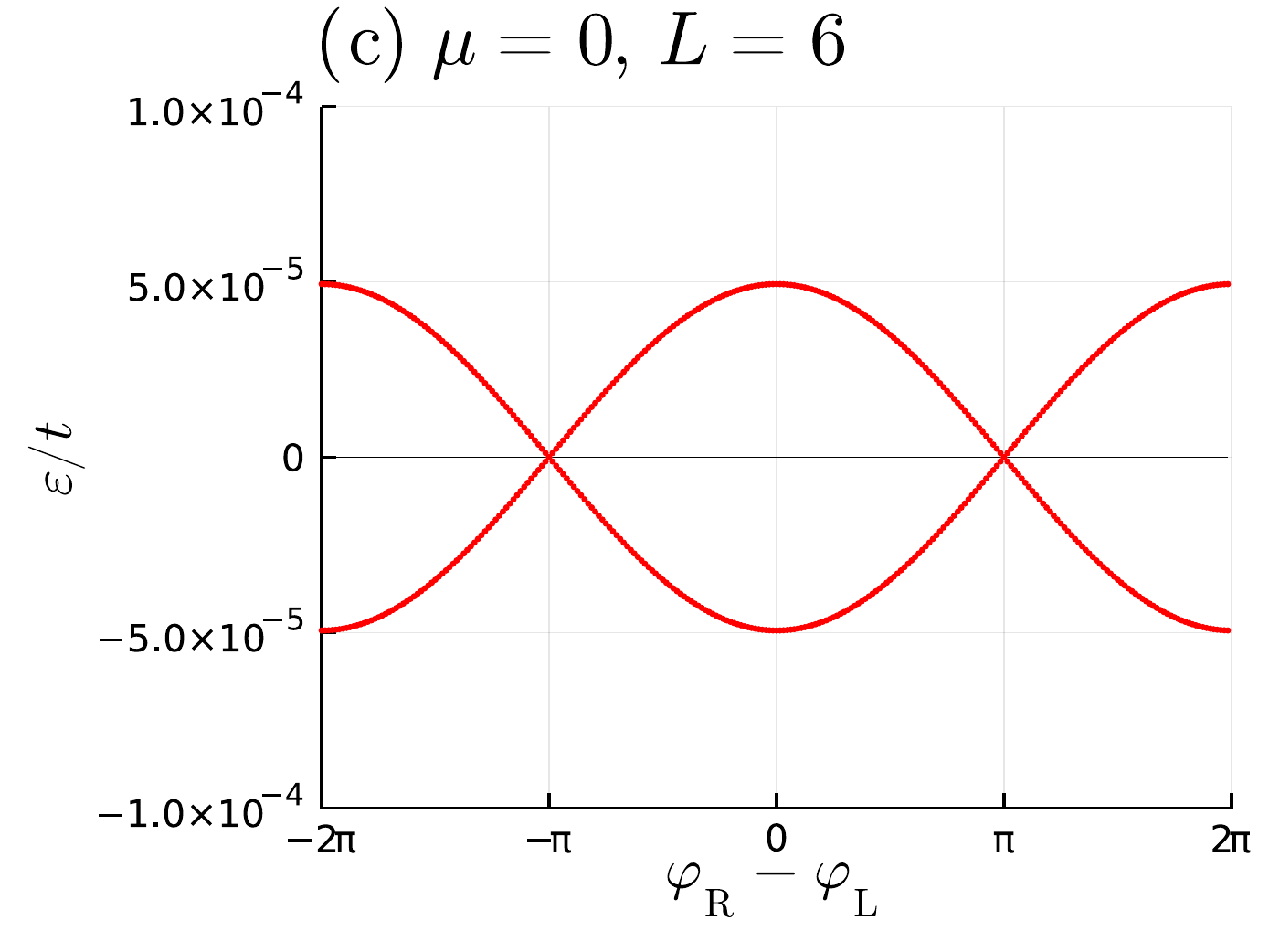}
 \caption{The $\varphi_{\text{R}} - \varphi_{\text{L}}$ dependence of the exact energy eigenvalues near zero energy, for (a) $\mu = 0, L = 5$, (b) $\mu = 0.001, L = 5$, and (c) $\mu = 0, L = 6$. The other parameters are $(\bar{\Delta}, \lambda) = (0.8, 0.01)$.}
 \label{fig:phidep_exact}
\end{figure}

\begin{figure}[tbp]
 \includegraphics[width=\linewidth]{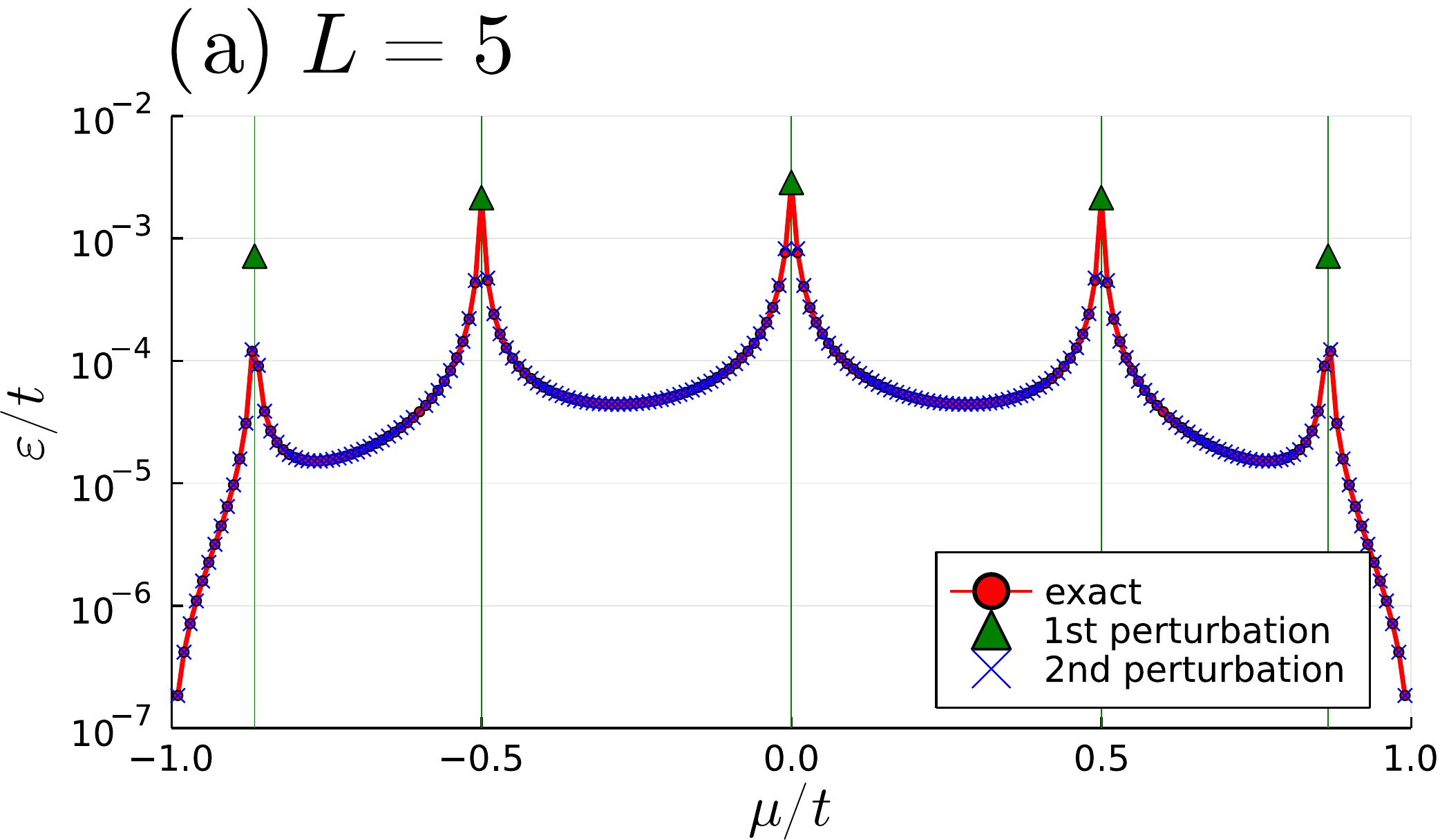}
 \includegraphics[width=\linewidth]{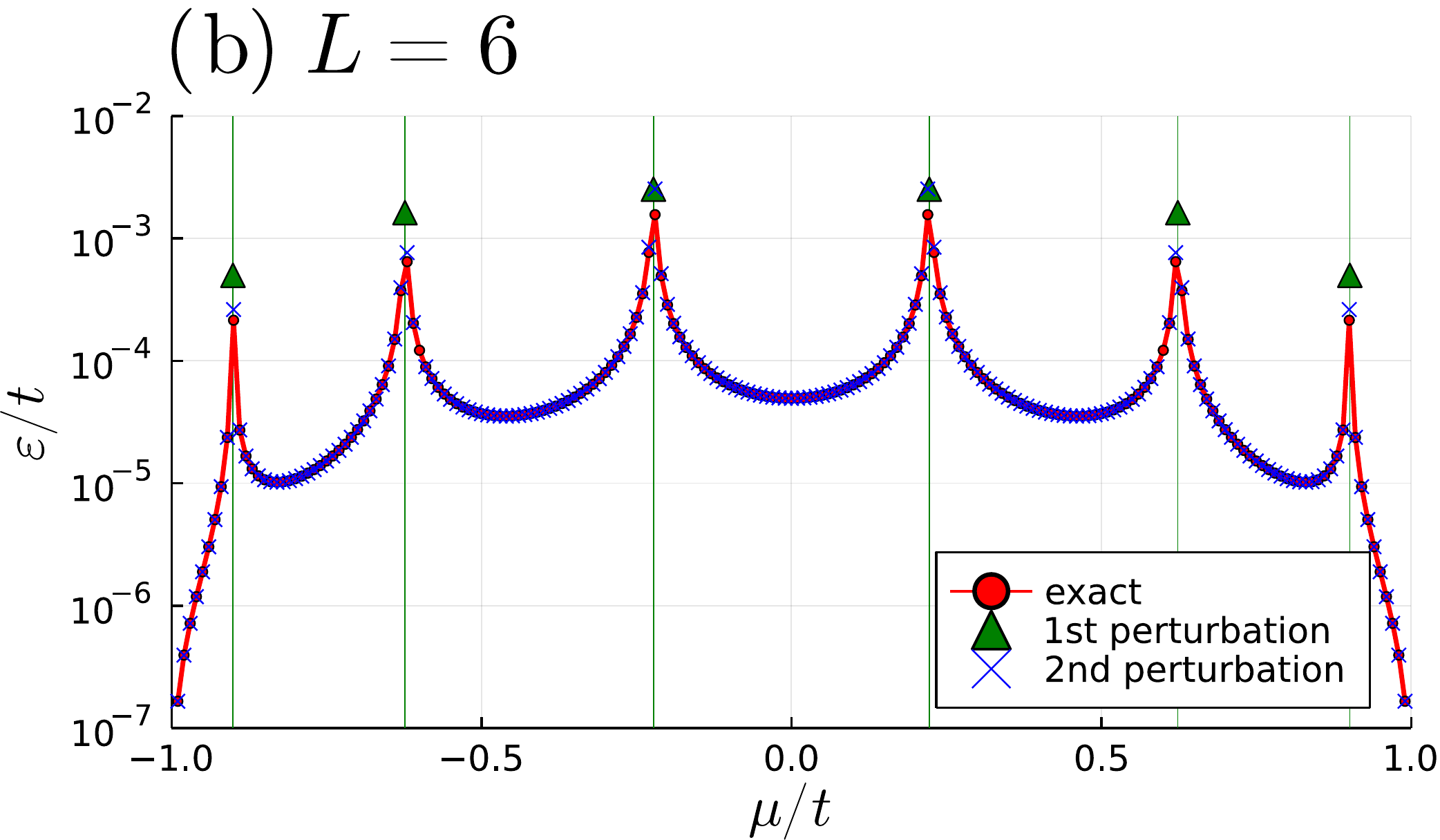}
 \caption{The chemical potential dependence of the smallest absolute eigenvalue for (a) $L = 5$ and (b) $L = 6$ (the red lines). The other parameters are $(\bar{\Delta}, \varphi_{\text{R}} - \varphi_{\text{L}}, \lambda) = (0.8, 0, 0.01)$. The green triangles (blue crosses) represent the results of the first-order (second-order) perturbation theory. $\varepsilon^{\text{N}}(q) = 0 \ (q = 1, \dots, L)$ is satisfied on the green lines.}
 \label{fig:mudep_exact}
\end{figure}

The magnitude of the zero-mode splitting corresponds to that of the Josephson effect, as discussed in Sec.~\ref{sec:effective_model}.
Therefore, the numerical exact calculation of energy levels gives us quantitative estimate of the Josephson current for arbitrary $\mu$ and $L$, i.e., for any parameter regime including both on- and off-resonant cases.
In Fig.~\ref{fig:mudep_exact} we show the minimum absolute value of the single-particle eigenenergies, as a function of the scaled chemical potential $\bar{\mu}$, fixing the relative phase $\varphi_{\text{R}} - \varphi_{\text{L}}$ to zero.
Obviously, the exact numerical results (the red lines in Fig.~\ref{fig:mudep_exact}) quantitatively agree with the results of the perturbation theory (the green and blue markers) for small $\lambda$~%
\footnote{We also confirmed that the perturbation theory deviates from the exact results at larger values of $\lambda$.}.
Therefore, the perturbation theory (Sec.~\ref{sec:perturbation}) and our effective model (Sec.~\ref{sec:effective_model}) help our understanding of the energy structures and the FP switches.

The numerically exact calculations confirm that the $4\pi$ periodic Josephson current is strongly enhanced at the resonance points satisfying $\varepsilon^{\text{N}}(q) = 0$ ($q = 1, \dots, L$).
The number of the enhanced peaks is of course determined by that of the on-resonance points in the parameter space, namely $L$.
In this sense, the lattice property of the junction is reflected in the magnitude of the fractional Josephson current.

\section{Summary and discussion}
\label{sec:summary}
In this paper, we investigated the 1D tight-binding model of the topological SNS junction where two semi-infinite Kitaev chains (superconducting electrodes) are weakly connected through the N region of $L$ sites.
We derived the energy level structures of the model by using a second-order (first-order) perturbation theory, when the eigenstates in the N region are off (on) resonance with MZMs.
Then we analyzed the effective four-state model interpolating the on- and off-resonance cases, and showed the existence of the fractional Josephson effect, where the Josephson current has a $4\pi$ periodicity as a function of the phase difference between the two superconductors~\cite{Kitaev2001, Kwon2004, Fu2009, Lutchyn2010, KTLaw2011, Nogueira2012, Pikulin2012, Beenakker2013_PRL, Zhang2014_PRB, Affleck2014, Sato2016, Sau2017}, in both cases.
In particular, resonant enhancement of the Josephson current is shown to occur.
Furthermore, we elucidated that the fractional Josephson current is proportional to the total FP involving occupied N levels and two MZMs.
Therefore, contrary to the naive expectation discussed in Introduction, the direction of the supercurrent flow is not reversed even when the chemical potential is changed, as long as the total FP is conserved.

Next, we numerically solved the original model using the exact diagonalization method of a corner-modified banded block-Toeplitz matrix~\cite{Alase2016, Cobanera2017, Alase2017, Cobanera2018}.
The calculation results are well compatible with the perturbation theory.
Furthermore, we showed rapid increase of the energy splitting of the MZMs in the on-resonance regime.
In other words, our results confirm the drastic enhancement of the fractional Josephson current to occur whenever a discrete N level and the MZMs are in resonance, which may be observable as a sizable signal for MZMs.
We expect that the resonant enhancement should also appear in the fractional ac Josephson current, which can be observed by the measurement of Shapiro steps~\cite{Rokhinson2012, Wiedenmann2016, Bocquillon2016, Li2018, Yu2018, Wang2018}.
In experiments it is necessary to distinguish the $4\pi$-periodic contribution to the Josephson current from the conventional $2\pi$-periodic one.
Therefore, our suggestion of the resonantly enhanced $4\pi$-periodic Josephson effect would be useful for the experimental detection.

Finally, we consider possible applications of our study as future works.
An interesting direction to explore is interaction effects.
On the one hand, in a weakly interacting case, the results of this paper will be qualitatively unchanged.
Indeed, the interacting Kitaev chain has a topologically nontrivial phase that can be continuously deformed to a topological phase in the noninteracting Kitaev chain without gap closing~\cite{Katsura2015}.
Furthermore, the repulsive interactions in the N region would not affect the current--phase relation in Eqs.~\eqref{eq:Josephson_current_off-resonance 2} and \eqref{eq:Josephson_current_on-resonance} of the noninteracting case, whereas its power-law dependence on the length $L$ is modified~\cite{Affleck2014}.
On the other hand, it may be possible to discuss a $\mathbb{Z}_4$ fractional Josephson effect if strong many-body interactions are taken into account~\cite{Zhang2014_PRL, Klinovaja2014, Orth2015, Vinkler-Aviv2017, Pedder2017, Shen2021}.
As for a technical aspect, our calculation method may be applicable to the study of topological physics of multi-terminal Josephson junctions, which has been discussed in recent studies~\cite{vanHeck2014, Yokoyama2015, Riwar2016, Eriksson2017, Meyer2017, Xie2017, Xie2018, Xie2019, Erdmanis2018, Huang2019, Repin2019, Kotetes2019, Nowak2019, Peralta2019, Kornich2019, Houzet2019, Pankratova2020}.

\begin{acknowledgments}
 The authors are grateful to Shingo Kobayashi and Youichi Yanase for fruitful discussions and comments.
 This work was supported by JST CREST Grant No. JPMJCR19T2.
\end{acknowledgments}

\appendix
\section{Exact solutions without tunneling Hamiltonian}
\label{app:exact_sol}
In this Appendix, we solve the eigenvalue equation of the Hamiltonian \eqref{eq:Hamiltonian} for $\lambda = 0$.
In this case, we can easily derive eigenenergies and eigenfunctions in the two S and one N regions, since they are completely decoupled from each other.
The eigenvalue problem can be recast as diagonalization of corner-modified banded block-Toeplitz matrix~\cite{Alase2016, Cobanera2017, Alase2017, Cobanera2018}.

\subsection{Superconducting regions: \texorpdfstring{$H_{\text{SL}}$}{HSL} and \texorpdfstring{$H_{\text{SR}}$}{HSR}}
First, we consider the two semi-infinite Kitaev chains located on either side of the SNS junction.

\subsubsection{Bulk solutions}
In order to construct bulk solutions, we need to consider a \textit{reduced bulk Hamiltonian} defined by~\cite{Alase2016, Cobanera2017, Alase2017, Cobanera2018}
\begin{align}
 H^{\text{S}}(z; \varphi) &= h_0^{\text{S}} + \left(z h_1^{\text{S}}(\varphi) + z^{-1} h_1^{\text{S}}(\varphi)^\dagger\right) \notag \\
 &= \frac{1}{2}
 \begin{bmatrix}
  -2\mu - t(z + z^{-1}) & \Delta e^{i\varphi} (z - z^{-1}) \\
  -\Delta e^{-i\varphi} (z - z^{-1}) & 2\mu + t(z + z^{-1})
 \end{bmatrix}.
\end{align}
Now a certain energy eigenvalue $\varepsilon$ is fixed as a given parameter.
Then an equation we should solve is represented as follows,
\begin{equation}
 z^2 \det\left(H^{\text{S}}(z; \varphi) - \varepsilon \bm{1}_2 \right) = 0.
 \label{eq:eigenz_sc}
\end{equation}
Solving the equation for $z$, we obtain four extended solutions,
\begin{equation}
 z = - w^{\text{S}}_l(\varepsilon) + \sigma \sqrt{\left[w^{\text{S}}_l(\varepsilon)\right]^2 - 1} \quad (l = 1, 2; \sigma = \pm),
 \label{eq:z_sc}
\end{equation}
where
\begin{equation}
 w^{\text{S}}_l(\varepsilon) = \frac{\bar{\mu} + (-1)^{l-1} \sqrt{\bar{\mu}^2 - (1 - \bar{\Delta}^2) (\bar{\mu}^2 + \bar{\Delta}^2 - \bar{\varepsilon}^2)}}{1 - \bar{\Delta}^2},
\end{equation}
and the characters with a bar represent quantities scaled by the hopping parameter: $\bar{\mu} := \mu / t$, $\bar{\Delta} := \Delta / t$, and $\bar{\varepsilon} := \varepsilon / t$.
Furthermore, eigenvectors for internal degrees of freedom corresponding to the four solutions are given by
\begin{equation}
 \begin{bmatrix}
  \bar{\Delta} e^{i\varphi} \sqrt{(w^{\text{S}}_l(\varepsilon))^2 - 1} \\
  \sigma (\bar{\varepsilon} + \bar{\mu} - w^{\text{S}}_l(\varepsilon))
 \end{bmatrix}
 \quad (l = 1, 2; \sigma = \pm).
 \label{eq:u_sc}
\end{equation}
For convenience, we label the roots~\eqref{eq:z_sc} and the corresponding eigenvectors~\eqref{eq:u_sc} as
\begin{alignat}{4}
 & z^{\text{S}}_1(\varepsilon), & \, & z^{\text{S}}_2(\varepsilon), & \, & \left[z^{\text{S}}_2(\varepsilon)\right]^{-1}, & \, & \left[z^{\text{S}}_1(\varepsilon)\right]^{-1},
 \label{eq:z_sc_label} \\
 & \Ket{u^{\text{S}}_{1+}(\varepsilon; \varphi)}, & & \Ket{u^{\text{S}}_{2+}(\varepsilon; \varphi)}, & & \Ket{u^{\text{S}}_{2-}(\varepsilon; \varphi)}, & & \Ket{u^{\text{S}}_{1-}(\varepsilon; \varphi)},
 \label{eq:u_sc_label}
\end{alignat}
where the indices are determined such that
\begin{equation}
 |z^{\text{S}}_1(\varepsilon)| \leq |z^{\text{S}}_2(\varepsilon)| \leq 1 \leq |z^{\text{S}}_2(\varepsilon)|^{-1} \leq |z^{\text{S}}_1(\varepsilon)|^{-1}.
 \label{eq:z_sc_inequality}
\end{equation}

In particular, we discuss MZMs localized at the ends of the superconductors, considering zero energy $\varepsilon = 0$, for which the solutions of Eq.~\eqref{eq:eigenz_sc} are given by
\begin{align}
 z &= \frac{- \mu \pm \sqrt{\mu^2 + \Delta^2 - t^2}}{t \pm \Delta}, \frac{- \mu \mp \sqrt{\mu^2 + \Delta^2 - t^2}}{t \pm \Delta} \notag \\
 &=: (\mathsf{z}^{\text{S}}_1)^{\pm 1}, (\mathsf{z}^{\text{S}}_2)^{\pm 1}
\end{align}
and the corresponding eigenvectors,
\begin{align}
 \Ket{\mathsf{u}^{\text{S}}_+(\varphi)} &:=
 \begin{bmatrix}
  i e^{i\varphi/2} \\ -i e^{-i\varphi/2}
 \end{bmatrix}
 & \text{for} \ z &= \mathsf{z}^{\text{S}}_{1, 2}, \notag \\
 \Ket{\mathsf{u}^{\text{S}}_-(\varphi)} &:=
 \begin{bmatrix}
  e^{i\varphi/2} \\ e^{-i\varphi/2}
 \end{bmatrix}
 & \text{for} \ z &= (\mathsf{z}^{\text{S}}_{1, 2})^{-1},
\end{align}
which are independent of $l = 1, 2$.
Supposing that $t, \Delta > 0$, we easily prove the following inequalities:
\begin{equation}
 \renewcommand{\arraystretch}{1.2}
 \begin{array}{ccc} \hline\hline
  \mu < -t & -t < \mu < t & t < \mu \\ \hline
  |\mathsf{z}^{\text{S}}_1| > 1 & |\mathsf{z}^{\text{S}}_1| < 1 & |\mathsf{z}^{\text{S}}_1| < 1 \\
  |\mathsf{z}^{\text{S}}_2| < 1 & |\mathsf{z}^{\text{S}}_2| < 1 & |\mathsf{z}^{\text{S}}_2| > 1 \\ \hline\hline
 \end{array}
 \label{eq:inequalities_zS}
\end{equation}

For simplicity, we assume that $\mathsf{z}^{\text{S}}_1 \neq \mathsf{z}^{\text{S}}_2$, namely $\mu^2 + \Delta^2 \neq t^2$.
Therefore, wavefunctions of the MZMs in the left and right superconducting regions are given by superpositions of the following eigenvectors,
\begin{align}
 \Ket{\psi^{\text{SL}}_{l \sigma}} &= \sum_{j=-\infty}^{0} (\mathsf{z}^{\text{S}}_l)^{\sigma j} \ket{j} \Ket{\mathsf{u}^{\text{S}}_\sigma(\varphi_{\text{L}})}, \\
 \Ket{\psi^{\text{SR}}_{l \sigma}} &= \sum_{j=L+1}^{\infty} (\mathsf{z}^{\text{S}}_l)^{\sigma j} \ket{j} \Ket{\mathsf{u}^{\text{S}}_\sigma(\varphi_{\text{R}})},
\end{align}
where $l = 1, 2$ and $\sigma = \pm$.

\subsubsection{Boundary matrix}
Next, we consider boundary conditions.
First, let us discuss the left superconducting chain ($-\infty < j \leq 0$).
According to Eq.~\eqref{eq:inequalities_zS}, two of the four eigenvectors $\Ket{\psi^{\text{SL}}_{l \sigma}}$ contribute to the Majorana wavefunction such that the wavefunction decays exponentially in the bulk superconductor.
Therefore, the open boundary condition at $j = 0$ is represented by a $2 \times 2$ boundary matrix,
\begin{align}
 & B^{\text{SL}}_{\lambda = 0}(\varepsilon = 0) \notag \\
 &=
 \begin{cases}
  \begin{bmatrix}
   \Braket{0 | H_{\text{SL}} | \psi^{\text{SL}}_{1+}} & \Braket{0 | H_{\text{SL}} | \psi^{\text{SL}}_{2-}}
  \end{bmatrix}
  & \mu < -t, \\[2mm]
  \begin{bmatrix}
   \Braket{0 | H_{\text{SL}} | \psi^{\text{SL}}_{1-}} & \Braket{0 | H_{\text{SL}} | \psi^{\text{SL}}_{2-}}
  \end{bmatrix}
  & -t < \mu < t, \\[2mm]
  \begin{bmatrix}
   \Braket{0 | H_{\text{SL}} | \psi^{\text{SL}}_{1-}} & \Braket{0 | H_{\text{SL}} | \psi^{\text{SL}}_{2+}}
  \end{bmatrix}
  & t < \mu.
 \end{cases}
\end{align}
Since $\Ket{\mathsf{u}^{\text{S}}_+(\varphi)}$ and $\Ket{\mathsf{u}^{\text{S}}_-(\varphi)}$ are linearly independent, the boundary matrix has a nontrivial kernel only when $-t < \mu < t$.
In this case, the boundary matrix is
\begin{equation}
 B^{\text{SL}}_{\lambda = 0}(0) = \frac{t - \Delta}{2}
 \begin{bmatrix}
  (\mathsf{z}^{\text{S}}_1)^{-1} e^{i\varphi_{\text{L}}/2} & (\mathsf{z}^{\text{S}}_2)^{-1} e^{i\varphi_{\text{L}}/2} \\[2mm]
  - (\mathsf{z}^{\text{S}}_1)^{-1} e^{-i\varphi_{\text{L}}/2} & - (\mathsf{z}^{\text{S}}_2)^{-1} e^{-i\varphi_{\text{L}}/2}
 \end{bmatrix},
\end{equation}
whose kernel is given by $
\begin{bmatrix}
 \mathsf{z}^{\text{S}}_1 & - \mathsf{z}^{\text{S}}_2
\end{bmatrix}^{\text{T}}
$.
As a result, the wavefunction of the MZM in the left superconducting region is
\begin{equation}
 \Ket{\psi^{\text{SL}}} = N^{\text{S}} \sum_{j=-\infty}^{0} \left[ (\mathsf{z}^{\text{S}}_1)^{-j+1} - (\mathsf{z}^{\text{S}}_2)^{-j+1} \right] \ket{j} \Ket{\mathsf{u}^{\text{S}}_-(\varphi_{\text{L}})},
\end{equation}
where $N^{\text{S}}$ is a normalization constant,
\begin{equation}
 N^{\text{S}} = \left[ 2 \sum_{j=1}^{\infty} \left| (\mathsf{z}^{\text{S}}_1)^j - (\mathsf{z}^{\text{S}}_2)^j \right|^2 \right]^{-1/2}.
 \label{eq:norm_S}
\end{equation}

For the right superconducting region, on the other hand, it is necessary to consider boundary conditions at $j \to \infty$ and $j = L + 1$.
By applying a similar discussion to the above, we can derive the right zero mode wavefunction:
\begin{equation}
 \Ket{\psi^{\text{SR}}} = N^{\text{S}} \sum_{j=L+1}^{\infty} \left[ (\mathsf{z}^{\text{S}}_1)^{j-L} - (\mathsf{z}^{\text{S}}_2)^{j-L} \right] \ket{j} \Ket{\mathsf{u}^{\text{S}}_+(\varphi_{\text{R}})},
\end{equation}
for $|\mu| < t$.

Note that, if $\mu^2 + \Delta^2 = t^2$, the above superpositions are not suitable for the zero modes, because the roots $z = (\mathsf{z}^{\text{S}}_1)^{\pm 1}$ are degenerate with $z = (\mathsf{z}^{\text{S}}_2)^{\pm 1}$.
Even in the case, we can write down an exact Majorana wavefunction by taking into account an eigenvector with a \textit{power-law prefactor}~\cite{Cobanera2017, Alase2017, Cobanera2018}:
\begin{align}
 \Ket{\psi^{\text{SL}}} &= 2 \tilde{N}^{\text{S}} \sum_{j=-\infty}^{0} (1 - j) (\mathsf{z}^{\text{S}}_1)^{-j} \ket{j} \Ket{\mathsf{u}^{\text{S}}_-(\varphi_{\text{L}})}, \label{eq:Kitaev_power-law_SL} \\
 \Ket{\psi^{\text{SR}}} &= 2 \tilde{N}^{\text{S}} \sum_{j=L+1}^{\infty} (j - L) (\mathsf{z}^{\text{S}}_1)^{j-(L+1)} \ket{j} \Ket{\mathsf{u}^{\text{S}}_+(\varphi_{\text{R}})}, \label{eq:Kitaev_power-law_SR}
\end{align}
where
\begin{equation}
 \tilde{N}^{\text{S}} = \left[ 8 \sum_{j=1}^{\infty} j^2 (\mathsf{z}^{\text{S}}_1)^{2(j - 1)} \right]^{-1/2}.
\end{equation}
For the detailed derivation of Eqs.~\eqref{eq:Kitaev_power-law_SL} and \eqref{eq:Kitaev_power-law_SR}, see Refs.~\cite{Cobanera2017, Alase2017}.

\subsection{A normal-metal region: \texorpdfstring{$H_{\text{N}}$}{HN}}
Now let us move on to solutions for the N region.

\subsubsection{Bulk solutions}
A reduced bulk Hamiltonian for the normal metal is given by
\begin{align}
 H^{\text{N}}(z) &= h_0^{\text{N}} + \left(z h_1^{\text{N}} + z^{-1} h_1^{\text{N} \dagger}\right) \notag \\
 &= \frac{1}{2}
 \begin{bmatrix}
  -2\mu - t(z + z^{-1}) & 0 \\
  0 & 2\mu + t(z + z^{-1})
 \end{bmatrix}.
\end{align}
Then we consider an eigenvalue equation at an energy $\varepsilon$:
\begin{equation}
 z^2 \det\left(H^{\text{N}}(z) - \varepsilon \bm{1}_2 \right) = 0.
\end{equation}
Solving the equation for $z$, we obtain four extended solutions,
\begin{align}
 z &= \frac{- (\mu + \varepsilon) \pm \sqrt{(\mu + \varepsilon)^2 - t^2}}{t}, \frac{- (\mu - \varepsilon) \pm \sqrt{(\mu - \varepsilon)^2 - t^2}}{t} \notag \\
 &=: \left[z^{\text{N}}_1(\varepsilon)\right]^{\pm 1}, \left[z^{\text{N}}_2(\varepsilon)\right]^{\pm 1},
 \label{eq:z_n}
\end{align}
and the corresponding following eigenvectors,
\begin{align}
 \Ket{u^{\text{N}}_1} &:=
 \begin{bmatrix}
  1 \\ 0
 \end{bmatrix}
 & \text{for} \ z = (z_1^{\text{N}})^{\pm 1}, \notag \\
 \Ket{u^{\text{N}}_2} &:=
 \begin{bmatrix}
  0 \\ 1
 \end{bmatrix}
 & \text{for} \ z = (z_2^{\text{N}})^{\pm 1},
 \label{eq:u_n}
\end{align}
which are independent of the energy $\varepsilon$.
Therefore, a wavefunction in the N region is given by superpositions of the following eigenvectors,
\begin{equation}
 \Ket{\psi^{\text{N}}_{l \sigma}(\varepsilon)} = \sum_{j=1}^{L} \left[z^{\text{N}}_l(\varepsilon)\right]^{\sigma j} \ket{j} \Ket{u^{\text{N}}_l},
\end{equation}
where $l = 1, 2$ and $\sigma = \pm$.

\subsubsection{Boundary matrix}
We consider (open) boundary conditions at $j = 1$ and $j = L$.
For the purpose, let us introduce a $4 \times 4$ boundary matrix,
\begin{widetext}
 \begin{align}
  B^{\text{N}}_{\lambda = 0}(\varepsilon) &=
  \begin{bmatrix}
   \Braket{1 | (H_{\text{N}} - \varepsilon \bm{1}) | \psi^{\text{N}}_{1+}(\varepsilon)} &
   \Braket{1 | (H_{\text{N}} - \varepsilon \bm{1}) | \psi^{\text{N}}_{1-}(\varepsilon)} &
   \Braket{1 | (H_{\text{N}} - \varepsilon \bm{1}) | \psi^{\text{N}}_{2+}(\varepsilon)} &
   \Braket{1 | (H_{\text{N}} - \varepsilon \bm{1}) | \psi^{\text{N}}_{2-}(\varepsilon)} \\[2mm]
   \Braket{L | (H_{\text{N}} - \varepsilon \bm{1}) | \psi^{\text{N}}_{1+}(\varepsilon)} &
   \Braket{L | (H_{\text{N}} - \varepsilon \bm{1}) | \psi^{\text{N}}_{1-}(\varepsilon)} &
   \Braket{L | (H_{\text{N}} - \varepsilon \bm{1}) | \psi^{\text{N}}_{2+}(\varepsilon)} &
   \Braket{L | (H_{\text{N}} - \varepsilon \bm{1}) | \psi^{\text{N}}_{2-}(\varepsilon)}
  \end{bmatrix} \notag \displaybreak[2] \\
  &= \frac{t}{2}
  \begin{bmatrix}
   1 & 1 & 0 & 0 \\[2mm]
   0 & 0 & -1 & -1 \\[2mm]
   (z^{\text{N}}_1(\varepsilon))^{L+1} & (z^{\text{N}}_1(\varepsilon))^{-(L+1)} & 0 & 0 \\[2mm]
   0 & 0 & -(z^{\text{N}}_2(\varepsilon))^{L+1} & -(z^{\text{N}}_2(\varepsilon))^{-(L+1)}
  \end{bmatrix}.
 \end{align}
\end{widetext}
The above matrix has nontrivial kernels $
\begin{bmatrix}
 1 & -1 & 0 & 0
\end{bmatrix}^{\text{T}}
$ and $
\begin{bmatrix}
 0 & 0 & 1 & -1
\end{bmatrix}^{\text{T}}
$ when
\begin{equation}
 \left[z^{\text{N}}_l(\varepsilon)\right]^{L+1} = \left[z^{\text{N}}_l(\varepsilon)\right]^{-(L+1)}, 
\end{equation}
therefore
\begin{equation}
 z^{\text{N}}_l(\varepsilon) = \exp\left(\frac{i\pi q}{L+1}\right) \quad (q = -(L+1), \dots, L),
 \label{eq:zN_nonperturbed}
\end{equation}
for $l = 1$ and $l = 2$, respectively.
Substituting Eq.~\eqref{eq:zN_nonperturbed} into the eigenvalue equation, we obtain the eigenenergies
\begin{equation}
 \varepsilon = \mp\left[\mu + t \cos\left(\frac{\pi q}{L+1}\right)\right] =: \pm \varepsilon^{\text{N}}(q) \quad \text{for} \ l = 1, 2.
\end{equation}
The normal wavefunctions for $l = 1, 2$ are represented by
\begin{equation}
 \Ket{\psi^{\text{N}}_l(q)} = N^{\text{N}} \sum_{j=1}^{L} \sin\left(\frac{\pi q j}{L+1}\right) \ket{j} \Ket{u^{\text{N}}_l},
\end{equation}
where $N^{\text{N}}$ is a normalization constant,
\begin{equation}
 N^{\text{N}} = \left[ \sum_{j=1}^{L} \sin^2\left(\frac{\pi q j}{L+1}\right) \right]^{-1/2} = \left( \frac{L+1}{2} \right)^{-1/2}.
 \label{eq:norm_N}
\end{equation}
Now we note that the above function represents the same state for $+q$ and $-q$.
Furthermore, the states for $q = 0, -(L+1)$ have no physical meaning since $\Ket{\psi^{\text{N}}_l(0)} = \Ket{\psi^{\text{N}}_l\bigl(-(L+1)\bigr)} = 0$.
Therefore, the domain of the integer $q$ should be restricted to $q = 1, \dots, L$.
As a result, we obtain $2N$ independent wavefunctions $\Ket{\psi^{\text{N}}_l(q)}$ for $l = 1, 2$ and $q = 1, \dots, L$, with eigenenergies $\pm \varepsilon^{\text{N}}(q)$.

%

\end{document}